\documentclass[10pt]{article}
\pdfoutput=1
\usepackage{jheppub,amsmath,amssymb}
\usepackage{enumerate}
\usepackage{bm}
\usepackage{bbm}
\usepackage{enumitem}
\usepackage{mathtools}
\usepackage[usenames,dvipsnames]{xcolor}
\usepackage{amsmath}
\usepackage{amsfonts}
\usepackage{amssymb}
\usepackage{graphicx}
\usepackage{hhline}
\usepackage{subfig}
\usepackage{hyperref}
\usepackage{color}
\usepackage{verbatim}
\usepackage{amsmath,amsthm,amssymb}
\usepackage{lscape}
\usepackage{float}
\usepackage{caption}
\usepackage{lscape}
\usepackage{breqn}
\usepackage{multicol}
\usepackage{makecell}
\usepackage{graphics}
\usepackage{tikz,todonotes}
\usepackage[utf8]{inputenc}
\usepackage{autobreak}
\usepackage{verbatim}
\usepackage{booktabs}
\usetikzlibrary{arrows,positioning,decorations.markings,decorations.pathmorphing,calc}
\pgfdeclarelayer{edgelayer}
\pgfdeclarelayer{nodelayer}
\usetikzlibrary{decorations.pathreplacing}
\pgfsetlayers{edgelayer,nodelayer,main}
\tikzset{none/.style={draw=none}}
\tikzset{new edge style 2/.style={black}}
\tikzset{new style 0/.style={black}}
\tikzset{rednode/.style={draw=none, scale=0.3pt,fill=red,circle, draw}}
\tikzset{redline/.style={line width=0.3mm,red}}
\tikzset{greyE/.style={line width=0.1mm,gray}}
\usepackage[utf8]{inputenc}
\usetikzlibrary{arrows,positioning,decorations.markings,decorations.pathmorphing,calc}

\definecolor{hyperref}{RGB}{026,028,087}

\newcommand{\beq}{\begin{equation}}
	\newcommand{\eeq}{\end{equation}}
\newcommand{\bal}{\begin{aligned}}
	\newcommand{\eal}{\end{aligned}}
\newcommand{\bea}{\begin{eqnarray}}
	\newcommand{\eea}{\end{eqnarray}}
\def\be{\begin{equation}}
	\def\ee{\end{equation}}

\def\beq{\begin{equation}}
	\def\eeq{\end{equation}}

\renewcommand{\[}{\left[}
\renewcommand{\]}{\right]}
\renewcommand{\L}{\mathcal L}

\def\be{\begin{equation}}
	\def\ee{\end{equation}}
\def\ba{\begin{eqnarray}}
	\def\ea{\end{eqnarray}}

\def\nn{\nonumber}

\def\d{\mathrm{d}}

\usepackage[normalem]{ulem}

\def\ba{\begin{eqnarray}}
	\def\ea{\end{eqnarray}}

\def\L{\mathcal{L}}

\def\d{\mathrm{d}}

\def\({\left(}
\def\){\right)}

\def\p{\partial}
\def\ie{{\em i.e. }}

\def\tilg8{\tilde{g}_8}

\begin{document}
	\preprint{Imperial/TP/2022/MC/05}	
	
	\title{Causal Effective Field Theories}

	\author{Mariana Carrillo Gonz\'alez, }
	\author{Claudia de Rham, }
	\author{Victor Pozsgay, }
	\author{Andrew J. Tolley}
	
	\affiliation{Theoretical Physics, Blackett Laboratory, Imperial College, London, SW7 2AZ, U.K.}
	
	\abstract{Physical principles such as unitarity, causality, and locality can constrain the space of consistent effective field theories (EFTs) by imposing two-sided bounds on the allowed values of Wilson coefficients. In this paper, we consider the bounds that arise from the requirement of low-energy causality alone, without appealing to any assumptions about UV physics. We focus on shift-symmetric theories, and consider bounds that arise from the propagation around both a homogeneous and a spherically-symmetric scalar field background. We find that low-energy causality, namely the requirement that there are no resolvable time advances within the regime of validity of the EFT, produces two-sided bounds in agreement with compact positivity constraints previously obtained  from $2 \rightarrow 2$ scattering amplitude dispersion relations using full crossing symmetry.}
	
	\maketitle
	
	\pagebreak

\section{Introduction}
From a bottom-up perspective, the construction of effective field theories (EFTs) based on symmetry principles allows us to compute observables in the infrared (IR) without the full knowledge of the ultraviolet (UV) completion of the theory. This has proven to be a useful approach not only in particle physics and cosmology but also when studying gravitational systems. While the EFT contains an infinite number of higher derivative interactions, at low energies only a finite number are relevant at a given order in the EFT expansion. Nevertheless, symmetry principles on their own are not sufficient to ensure that the EFT is unitary and causal. Imposing these physical principles leads to constraints on the possible values of the coefficients in the Wilsonian effective action of the low energy EFT \cite{deRham:2022hpx}. A well-known approach for bounding these Wilson coefficients consists of looking at dispersion relations for $2 \rightarrow 2$ scattering amplitudes and engineering positive bounded functions of the scattering amplitude \cite{Adams:2006sv,deRham:2017avq,deRham:2017zjm,Bellazzini:2015cra,Bellazzini:2016xrt,Vecchi:2007na,Nicolis:2009qm}\footnote{Earlier approaches in the chiral perturbation theory context are found in \cite{yndurain1972rigorous,PhysRevD.31.3027,Pennington:1994kc,Ananthanarayan:1994hf,Comellas:1995hq,Manohar:2008tc}.}. The associated positivity bounds require assumptions about the UV completion such as unitarity, locality, causality, Poincar\'e symmetry, and crossing symmetry. Additionally, one can obtain stronger bounds when considering weakly coupled tree-level UV completions. In recent years this has proven to be a fruitful approach (see for example \cite{Cheung:2016yqr,Bonifacio:2016wcb,deRham:2018qqo, Alberte:2019xfh, Alberte:2019zhd, Wang:2020xlt,deRham:2017xox,deRham:2017imi,Davighi:2021osh,Bellazzini:2019bzh,Sinha:2020win,Chandrasekaran:2018qmx,Wang:2020jxr,Du:2021byy,Li:2021lpe,Bellazzini:2020cot,Arkani-Hamed:2020blm,Guerrieri:2020bto,Tolley:2020gtv, Caron-Huot:2020cmc,Hebbar:2020ukp,Bellazzini:2021oaj,Chiang:2021ziz,Chiang:2022ltp}). Crucially in \cite{Tolley:2020gtv, Caron-Huot:2020cmc} it was shown that incorporating the constraints of full crossing symmetry, now referred to as null constraints, imposes two-sided positivity bounds generically on the space of all Wilson coefficients\footnote{This phenomenon was already noted in \cite{Cheung:2016yqr,Bonifacio:2016wcb,deRham:2018qqo,Alberte:2019xfh,Alberte:2019zhd,Wang:2020xlt} in the context of massive spin-1 and -2 theories where the two-sidedness comes from consideration of different external polarizations.}. The purpose of the present work is to show that these two-sided bounds can be largely anticipated from low-energy causality considerations alone. \\

The extension of positivity bounds to (massless) gravitational theories and arbitrary curved spacetimes, and more specifically to time-dependent gravitational backgrounds is not straightforward \cite{Alberte:2020jsk,Alberte:2020bdz}. Gravitational amplitudes in Minkowski spacetime have recently been incorporated by using dispersive arguments that evade the $t$-channel pole inevitable in gravitational amplitudes \cite{Caron-Huot:2022ugt,Bern:2021ppb,Chowdhury:2021ynh,Caron-Huot:2022jli,Chiang:2022jep} or account for it by its implied Regge behaviour \cite{Tokuda:2020mlf,Aoki:2021ckh,Noumi:2021uuv,Alberte:2021dnj,Herrero-Valea:2020wxz,Herrero-Valea:2022lfd}. While perturbative unitarity rules can be generalised on curved spacetime \cite{deRham:2017aoj}, analyticity has proven more challenging and some initial explorations of positivity bounds to curved spacetimes were proposed in \cite{Baumann:2015nta,Baumann:2019ghk,Afkhami-Jeddi:2018own}. Further analyses considering that the bounds arising from positivity constraints around a Minkowski vacuum can be translated into bounds for Wilson coefficients around a curved vacuum are examined in \cite{Melville:2019wyy,deRham:2021fpu,Traykova:2021hbr,Kim:2019wjo,Herrero-Valea:2019hde,Ye:2019oxx}. The main difficulties in constructing dispersion relations in curved backgrounds arise due to the broken Lorentz symmetries and the lack of an S-matrix. Some progress has been made recently for broken Lorentz boost theories \cite{Grall:2021xxm} and in de Sitter spacetimes where there is an equivalent notion of positivity of spectral densities \cite{Bros:1995js,Sleight:2019hfp,Sleight:2020obc,Sleight:2021plv,Hogervorst:2021uvp,DiPietro:2021sjt}. By contrast the causality approach discussed here is easily generalizable to curved spacetimes.\\

In this paper, we will focus on constraints arising purely from causality in the low energy regime.  Our central tool is the scattering time delay well studied in non-relativistic scattering \cite{Eisenbud,Wigner:1955zz,Smith:1960zza,Martin:1976iw,DECARVALHO200283} and gravitational scattering \cite{Shapiro:1964uw,Gao:2000ga} which describes in the semi-classical (WKB) or eikonal approximation the delay of a scattered wave relative to a freely propagating wave. Causality violation is associated with the presence of a resolvable time advance, and this criterion has in recent years been utilised to impose similar bounds on Wilson coefficients \cite{Camanho:2014apa,Camanho:2016opx,Bai:2016hui,Goon:2016une,Hinterbichler:2017qcl,Hinterbichler:2017qyt,AccettulliHuber:2020oou,Bellazzini:2021shn,Serra:2022pzl,Chen:2021bvg,deRham:2021bll} which, importantly, do not require any assumption on the UV behaviour of the theory.
Indeed, since small superluminalities could lead to correlation functions having support outside of the light-cone when present at large distances, the violations of causality can be measured within the low energy EFT that describes the infrared physics. In a generic EFT, the higher-derivative interactions will modify the equation of motion for the propagation of a perturbation around an arbitrary background rendering a sound speed $c_s\neq 1$. Note however that a small superluminal low-energy speed is not necessarily in contradiction with causality since the would-be observation detecting violations of causality could turn out to be unmeasurable within the regime of validity of the EFT \cite{Hollowood:2015elj,deRham:2019ctd,deRham:2020zyh}. \\

For a local field theory in Minkowski spacetime, causality tells us that the retarded Green's function evaluated in an arbitrary quantum state does not have support outside of the forward Minkowski light-cone. For a generic EFT, locally the propagation of information is encoded in effective metric arising in the hyperbolic equations of motion for small fluctuations around a given background which at leading order is determined by the sound speed $c_s$ and reads
\begin{equation}
	\mathrm{d} s_{\text{eff.}}^{2}=-c_{s}^{2}(x^\mu,\omega) \mathrm{d}t^{2}+\mathrm{d} \vec{x}^{2} \ . \label{eq:effg}
\end{equation}
Generically this speed is dependent on the momentum scale/frequency of propagating fluctuations $\omega$. Causality does not directly impose constraints on the phase velocity, but it requires that its high frequency (high $\omega$) limit, that is, the front velocity is luminal $c_{s}^{2}(x^\mu,\infty)=1$. This determines the support of the retarded propagator and implies that information propagates (sub)luminally. Furthermore, it can be shown that causality implies analyticity of the scattering amplitude and refractive index in the upper half complex $\omega$-plane \cite{Nussenzveig:1972tcd,HamiltonBook,PhysRev.104.1760}. \\

Here, we will only focus on the causal properties of the EFT as encoded on light-cones defined by the effective metric in Eq.~\eqref{eq:effg} in the low frequency regime where the EFT is under control. In the EFT, the true front velocity is unknown, as is whether there is a Lorentz invariant UV completion. Furthermore demanding locally the strict bounds $c_{s}^{2}(x^\mu,\omega) \le 1$ is too strong since the associated apparent superluminality may be unresolvable within the EFT (furthermore in the gravitational context the local speed is sensitive to field redefinitions, although this last subtlety will not be relevant here\footnote{On curved backgrounds, the notion of asymptotic causality \cite{Gao:2000ga,Camanho:2014apa} (requiring the absence of superluminalities as compared to the asymptotic flat metric which imposes bounds on the net scattering time delay) is a physical requirement, but it does not always capture the full implications of causality. In fact, it leads to weaker bounds than the notion of infrared causality \cite{Hollowood:2015elj,deRham:2019ctd,deRham:2020zyh,deRham:2021bll,Chen:2021bvg}, (requiring the absence of superluminalities as compared to the local metric which imposes bounds on the net scattering time delay minus the Shapiro time delay).}). The presence of local low energy superluminality does not in itself imply the possibility of creating closed time-like curves. For that these superluminalities ought to be maintained for sufficiently large regions of spacetime. A cleaner diagnostic is the scattering time delay which is defined from the S-matrix and is hence independent of field redefinitions.
The scattering time delay for a given incident state containing a particle of energy $\omega$ may be defined in terms of the $S$-matrix by
\be
\Delta T = -i \left\langle {\rm in} \right| \hat S^{\dagger} \frac{\p }{\p \omega} \hat S \left|{\rm  in} \right\rangle \, .
\ee
The scattering phase shifts may be defined as the eigenvalues of the $S$-matrix, $\hat S|{\rm in}\rangle=e^{2i\delta}|{\rm in}\rangle$, so that in an incident eigenstate the time delay is simply
\be
\Delta T =2 \frac{\p  \delta }{\p \omega}\, .
\ee
For example, for one-particle scattering in a spherically symmetric background, the $S$-matrix diagonalises in multipoles $\ell$ and we may define the associated multipole time delays
\be
\Delta T_{\ell} =2 \frac{\p  \delta_{\ell} }{\p \omega}\Big|_\ell \, .
\ee
In the large-$\ell$ limit, we may consider scattering at fixed impact parameter $b=(\ell+1/2)\omega^{-1}$, giving the time delay traditionally calculated in the Eikonal approximation \cite{Wallace:1973iu,Wallace:1973ni}
\be
\lim_{\ell \rightarrow \infty} \delta_{\ell=b \omega-1/2}(\omega) = \delta_{\rm Eikonal}(\omega,b) \, ,
\ee
for which the time delay is (see for example \cite{Camanho:2014apa})
\be
\Delta T_{b} =2 \frac{\p  \delta_{\ell} }{\p \omega} \Big|_b \, .
\ee

The signature of true causality violation would be the manifest existence of closed-time-like-curves within the regime of validity of the EFT, however it is understood that such phenomena are akin to experiencing a {\it resolvable}\footnote{Strict positivity of the scattering time delay is sometimes incorrectly imposed. This is not required since the time delay is only a meaningful indication of causality in the semi-classical region (WKB or eikonal).} scattering time advance, (within the regime of validity of the EFT). The resolvability requirement comes from the uncertainty principle which is reflected in the fact that a time advance no bigger than the uncertainty $\Delta t \sim\omega^{-1}$ is clearly not in conflict with causality.
Indeed in general, as is well understood, scattering time advances can be mildly negative without contradicting causality, but only in a bounded way. For example for s-wave (monopole) scattering in a spherically symmetric potential which vanishes for $r>a$, causality imposes the bound on the scattering time delay of the form \cite{Eisenbud,Wigner:1955zz,Smith:1960zza,Martin:1976iw,DECARVALHO200283}
\be
\Delta T_{\ell =0} \ge - \frac{2a }{v} + \frac{1}{k v}\sin(2 k a+\delta_0)  \ge - \frac{2a }{v} - \frac{1}{k v}\, ,
\ee
with $v$ the group velocity and $k$ the momentum with $\omega \sim \mathcal{O}(k v)$. The first term gives the allowed time advance associated with the spherical waves scattering of the boundary $r=a$, and the second term gives an allowed time advance due to the wave nature of propagation, i.e. the uncertainty principle. For the intermediate scale frequencies and smooth backgrounds considered in what follows the first term will be absent (see Appendix \ref{timeadvance} for a discussion) but we must still allow for the uncertainty principle. In other words, we will consider frequencies larger than the scale of variation of the background (within the WKB semi-classical region) and sufficiently high such that we do not encounter any potential barriers, but within the regime of validity of the EFT. All these conditions will be carefully monitored throughout the analysis performed below. Note that lower frequencies do not probe the support of the retarded Green's function and hence are not probing causality. Working in the regime of validity of the WKB approximation, our de facto relativistic causality requirement is that
\be
\label{defacto}
\Delta T  \gtrsim -\frac{1}{\omega} \, .
\ee
applied in the relativistic region where the background is sufficiently smooth and no potential barrier is encountered on scales set by the wavelength $\omega^{-1}$ such that the hard sphere type time advances $- 2a/v$ are absent.\\

The goal of this paper is then to determine constraints we obtain on a given EFT by imposing \eqref{defacto} around different backgrounds. Since our primary concern will be non-gravitational scalar field theories, we can choose to probe the EFT by adding an external source. This device allows us to consider backgrounds which are not solutions of the unsourced background equations of motion. By choosing different sources, we can adjust the background solution to probe different possible scattering phases, and by extremising over the choices of backgrounds we will be able to obtain competitive constraints from the scattering time delay.\\

The rest of the paper is structured as follows. In Section \ref{sec:speed}, we introduce the shift-symmetric low energy scalar EFT we will be considering and discuss the positivity constraints that arise from consideration of their scattering amplitudes. We also provide generic arguments for the expected time delay within a WKB approach on generic backgrounds. For concreteness, we then focus  on specific profiles for the rest of the manuscript. In Section \ref{sec:hom}, we consider  the simple case of a homogeneous background and argue for the need of less symmetric configurations to make further contact with positivity bounds. We then proceed to consider the scattering of perturbations around a spherically-symmetric background in Section \ref{sec:Spherical}. We examine two limits: one where the waves have no angular dependence and the other where they have large angular momentum. For each of these cases, we spell out carefully the conditions for the validity of the EFT and the WKB approximation. After computing the time delay and requiring that we cannot obtain a resolvable violation of causality we obtain bounds on the Wilson coefficients of the EFT. The case of no angular momentum gives rise to a lower bound while the large angular momentum case draws an upper bound that approaches the non-linear positivity bounds obtained in \cite{Tolley:2020gtv,Caron-Huot:2020cmc}. Lastly, we discuss our results and conclude in Section \ref{sec:Concl}. In the Appendices, we show details of our calculations at higher orders in the EFT and for large angular momentum. We also explain our setup for obtaining bounds on the Wilson coefficients.

\section{Low energy effective field theory and propagation speed} \label{sec:speed}

In this paper, we consider the requirements for a scalar effective field theory to be causal. For pedagogical simplicity we focus on theories invariant under a shift symmetry $\phi\rightarrow\phi+c$. Since we are interested in comparing the constraints arising from $2 \rightarrow 2$ tree-level scattering, we will consider only operators up to quartic order in the field $\phi$, and we will ensure to work in a regime where operators that are higher order in the field remain irrelevant to our causality considerations.
In the following, we work with a minimal set of such independent operators up to dimension-12, so that our Lagrangian is given by \cite{Solomon:2017nlh}
\begin{align}
	\L = - \frac12 (\p \phi)^2 - \frac12 m^2 \phi^2+ \frac{g_8}{\Lambda^4} (\p \phi)^4
	  + \frac{g_{10}}{\Lambda^6} (\p \phi)^2 \Big[ (\phi_{, \mu \nu})^2  -  (\Box \phi)^2 \Big] + \frac{g_{12}}{\Lambda^8} (( \phi_{, \mu \nu} )^2 )^2 - g_{\text{matter}} \phi J\, ,
	\label{eq:L}
\end{align}
where $(\phi_{,\mu \nu})^2= \p_{\mu} \p_{\nu} \phi \p^{\mu} \p^{\nu} \phi$, $(\p \phi)^2=\p_\mu\phi\p^\mu\phi$, $g_{\text{matter}}$ is the coupling strength to external matter and $J$ is an arbitrary external source. Note that for convenience we choose to write down the dimension-10 operator as the quartic Galileon\footnote{The time delay remains manifestly invariant under field redefinitions as long as we can neglect boundary terms. This can be seen for instance explicitly in Section \ref{sec:monopole} for the zero angular momentum case up to the EFT order that we consider here.} \cite{Nicolis:2008in}. The scale $\Lambda$ has been introduced as the standard cutoff of this low energy EFT. Note that even though some EFTs may be reorganised so as to remain valid beyond $\Lambda$ (see for instance \cite{deRham:2014wfa} for a discussion), here we take the more conservative approach and consider the low energy EFT to break down at $\Lambda$. Except when we consider the case $g_8=0$, it proves convenient to redefine $\Lambda$ so that $g_8=1$.

\paragraph{Positivity Bounds:}
The aim of what follows is to establish to which extent positivity bounds constraining $\{ g_8 , g_{10}, g_{12} \}$ can be reproduced using low energy infrared causality arguments (i.e. the statement of causality as manifested directly at the level of the low energy EFT without any prior knowledge on the embedding of this EFT within a unitary high energy completion).
Technically, the derivation of the positivity bounds requires the presence of a mass gap and to this purpose, in principle, we can always introduce a shift-symmetry-breaking mass term in Eq.~\eqref{eq:L}. The mass term can indeed be treated as an irrelevant deformation of the shift-invariant Lagrangian which, at the quantum level, does not induce any further symmetry-breaking operators \cite{Burrage:2010cu,deRham:2017imi}.
In the following, we will be working in the limit  $m\ll \omega$ where the mass term can be neglected (hence effectively restoring shift symmetry). The positivity bounds from \cite{Tolley:2020gtv,Caron-Huot:2020cmc} can be translated into bounds on the Wilson coefficients appearing in the Lagrangian of Eq.~\eqref{eq:L} by using Table \ref{tab:dictionary} and read
\begin{equation}
	g_8 > 0, \qquad g_{12} > 0, \qquad g_{10} < 2 g_8, \qquad g_{12} < 4 g_8, \qquad - \frac{16 }{3} \sqrt{g_8 g_{12}} < g_{10} < \sqrt{g_8 g_{12}} \,. \label{eq:pos}
\end{equation}
From the above, only the left hand side of the last bound is derived by using full crossing symmetry whereas the other bounds follow from standard fixed $t$ dispersion relations.

\paragraph{Causality:}
Violations of causality can occur when superluminal speeds can be consistently maintained within a region of spacetime so as to lead to a physical support of the retarded propagator outside the standard Minkowski light-cone. In this section, we will compute the low frequency propagation speed of a perturbation $\psi=\phi-\bar \phi$ living on an arbitrary background $\bar{\phi}$  created by an external source $J$. For this, we work within the WKB approximation such that the background's scale of variation, $r_0$, is much larger than the scale on which the perturbation varies ($\omega^{-1}$).  In Section~\ref{sec:hom} we will perform a precise analysis of the possible violations of causality arising in a homogenous background and in a static and spherically-symmetric background in Section~\ref{sec:Spherical}, but for now it instructive to consider perturbations given by a plane wave $\partial_\mu\psi=i k_\mu \psi$.

The equation of motion for the scalar field $\phi$ is given by
\begin{eqnarray} \label{eq:lag}
	\Box \phi &=&  \frac{4g_8}{\Lambda^4} \left(\phi_{,\mu} (\p \phi)^2 \right)^{,\mu}
	-  \frac{2g_{10}}{\Lambda^6} \left[(\Box \phi)^3- 3 \Box \phi (\phi_{,\mu \nu})^2 + 2 (\phi_{,\mu \nu})^3 \right]
	 -  \frac{4g_{12}}{\Lambda^8} \left( \phi_{, \alpha \beta} (\phi_{,\mu \nu})^2 \right)^{, \alpha \beta} \\
&-& g_{\text{matter}} J  \,.\nonumber
\end{eqnarray}
In the WKB approximation, we assume that perturbations can be characterised by plane waves with wave-vector $k_\mu=(\omega,\bf{k})$. In the regime of validity of the EFT, the $g_{8,10,12}$ operators considered in \eqref{eq:L} are treated perturbatively implying $-k_\mu k^\mu=\omega^2-|{\bf{k}}|^2=(c_s^2-1)|{\bf{k}}|^2\ll |{\bf{k}}|^2$. One should note that remaining within the regime of validity of the EFT requires
\begin{equation}
	\frac{\partial {\phi}}{\Lambda^2}\equiv\delta_1 \ll 1 \ ,\qquad  \frac{\partial^{p+1} {\phi}}{\Lambda^{p+2}}\equiv \delta_1 \, \delta_2^p \ll 1 \ , \label{eq:validity}
\end{equation}
where $p\in \mathbb{N}$ and the derivatives can hit the background or the perturbation. The most stringent bounds are obtained when $p\rightarrow \infty$. Since we are interested in contributions up to dimension-12 operators, we will consider the expansion up to order $\delta_1^2\delta_2^2,\delta_1^4$ and assume $\delta_1^2\ll\delta_2, \ \delta_2^2\ll\delta_1$. Thus, at order $\delta_1^2\delta_2^2,\delta_1^4$ and leading order in $\omega r_0$ we have:
\begin{align}
	c_s^2|{\bf{k}}|^2=&|{\bf{k}}|^2-g_8\frac{8}{\Lambda^4}(k^\mu\partial_\mu\bar{\phi})^2+g_8^2\frac{32}{\Lambda^8}(k^\mu\partial_\mu\bar{\phi})^2(\partial\bar{\phi})^2-g_{12}\frac{8}{\Lambda^8}(k^\mu k^\nu\partial_\mu\partial_\nu\bar{\phi})^2 \nonumber \\
	& +g_{10}\frac{12 k^\mu k^\nu}{\Lambda^6}\left(\partial_{\mu}\partial_\rho\bar{\phi}\partial^\rho\partial_\nu \bar{\phi}-\square\bar{\phi}\partial_\mu\partial_\nu\bar{\phi}\right) \ ,
\end{align}
where we can immediately see that the $g_8$ and $g_{12}$ contributions are sign definite which are directly equivalent to the first two positivity bounds included in \eqref{eq:pos}. This direct equivalence was pointed out for the $g_8$ operator in  \cite{Adams:2006sv}.
In what follows we shall attempt to make contact with the remaining bounds in \eqref{eq:pos} but note that the contributions of $g_8$ and $g_{12}$ to the speed imply that, within this framework  we consider here, it will be impossible to reproduce the bound $g_{12} < 4 g_8$  from pure infrared causality considerations since, in the absence of $g_{10}$, positivity of both $g_8$ and $g_{12}$ is sufficient to prevent causality violation in this limit.

To establish the basic setup, it will be useful to start by looking at the simple example of a homogeneous background first (even though no further bounds will be derived), before proceeding to a more instructive  spherically-symmetric situation which will allow us to make further contact with the remaining bounds of \eqref{eq:pos}.

\paragraph{Time Delay:}

To understand whether the perturbations are causal around an arbitrary background, we need to consider a hierarchy between the scales of variation of the background and the perturbations, namely, $\lambda_{\text{background}}\gg \lambda_{\text{perturbation}}$. Hence, we can use the WKB approximation to obtain the phase shift experienced by the scattered perturbation from which the time delay is easily computed. Considering the wave nature of the scattering and the uncertainty principle, we define a resolvable time advance as one that satisfies
\begin{equation}
	\omega	\Delta T  < - 1 \ , \label{eq:NoCausality}
\end{equation}
where $\omega$ is the asymptotic energy of the scattered state. This states that a resolvable time advance needs to be larger than the resolution scale of {\it{geometric optics}}\footnote{The geometric optics or eikonal limit assumes that the scattering problem can be described in terms of particle trajectories with large impact parameters and that the energies of the asymptotic states are large.}, \cite{Hollowood:2015elj,deRham:2020zyh}. If Eq.~\eqref{eq:NoCausality} is satisfied within the WKB approximation and the regime of validity of the EFT, then we have an observable violation of causality. At leading order in the EFT, the requirement in Eq.~\eqref{eq:NoCausality} can be equivalently written in terms of the scattering phase shift as $\delta^{\text{EFT}}<-1$. However, this does not hold when including higher EFT corrections that modify the speed of sound with $\omega$-dependent contributions. In such cases, one should simply consider the bound in Eq.~\eqref{eq:NoCausality}. A commonly taken approach to understanding causality bounds consists of working in the eikonal (geometric optics) limit. This amounts to considering the scattering of waves with large angular momentum $\ell$ so that the dynamics can be described in terms of the scattering of particle trajectories with fixed impact parameter $b=(\ell+1/2) \omega^{-1}$. We shall consider both this region and the small $\ell$ region which can also be described semi-classically. We will see that exploration of both limits is complementary when imposing bounds on Wilson coefficients from the requirement of causality in the EFT and will give rise to the two-sided bounds known from other considerations.

As already noted, demanding locally that $c_s(\omega) \le 1$ is too strict a requirement. The leading-order contributions to violations of causality, as encoded in the support of the retarded Green's function, are determined by the light-cones defined by the effective metric in the EFT equations of motion \cite{Caldwell:1993xw,deRham:2019ctd} which is given by Eq.~\eqref{eq:effg}. Thus, acausality can be measured by integrating the effects of the sound speed. More precisely, by measuring whether the scattered waves can propagate outside the Minkowski light-cone and if this effect is observable within the regime of validity of the EFT. It is clear that a subluminal speed will always lead to a causal theory. On the other hand, small (as dictated by the validity of the EFT) superluminalities do not violate causality if they do not have support in a large region of spacetime. 

In what follows we perform a careful treatment of causality for homogeneous backgrounds and static and spherically symmetric ones. Note that we always use the same definition of causality, but depending on the symmetries of the background it might be more natural to express the support outside the Minkowski lightcone as given by a timelike or a spatial observable. For example, when we work with a homogeneous background we have spatial momentum conservation and hence the natural observable is the spatial displacement of the lightcone defined by the effective metric. On the other hand, when we work with spherically symmetric backgrounds, we have energy conservation and it is more natural to describe the support outside the Minkowski lightcone with a timelike displacement. Both of these quantities encode the same physics and capture the support of the retarded Green's function; hence both encode the same causality criterion.

\section{Homogeneous background} \label{sec:hom}
In this section, we will derive the dispersion relation for a homogeneous background. To do so, we consider the equation of motion in Eq.~\eqref{eq:lag} and perturb the field around a homogeneous background $\bar{\phi}(t)$ which varies on a time scales of order $H^{-1}$, which we shall consider as being constant to a first approximation. To access the information encoded in the retarded Green's function, we consider a perturbative setup in which we derive a perturbative, second-order in time, hyperbolic, equation of motion for the perturbations. Any higher-order ($>2$) time derivative can be iteratively removed at each  order in the EFT expansion. Schematically, the equations of motion for the perturbation reads
\begin{equation}
	\ddot{\psi} + A \dot{\psi} + B \psi = 0 \,,
\end{equation}
where $A$ and $B$ are functions of the background $\bar{\phi}(t)$ and its derivatives, the wavenumber $k$, the coupling constants $g_I$, and the energy scale $\Lambda$. For convenience, the friction term can be removed by performing a field redefinition of the form $\psi(t) = f(\bar{\phi}(t)) \psi_0(t)$ leading to the perturbation equation
\begin{equation}
	\ddot{\psi}_0 + \left( B - \frac{A^2 + 2 \dot{A}}{4} \right) \psi_0 = 0 \,.
\end{equation}
This allows us to write down an effective dispersion relation for the perturbations as
\begin{equation}
	\omega^2 = m_{\rm eff}^2 + c_s^2( {\bf k}) |{\bf{k}}|^2 \,,
\end{equation}
where $m_{\rm eff}^2$ is the effective mass square and $c_s^2( {\bf k})$ is the ${\bf k}$-dependent square sound speed. Our definition of the sound speed corresponds to a momentum-dependent phase velocity. Note that at leading order in $|{\bf{k}}|$, the notions of phase velocity and group velocity are equivalent when the mass is negligible, which is the case under consideration. Considering only the $g_8, g_{10},$ and $g_{12}$ operators, we find at order $\delta_1^2\delta_2^2$, (where the expansion parameters $\delta_1$ and $\delta_2$ are defined in \eqref{eq:validity}),
\begin{align}
	m_{\rm eff}^2&=-\frac{12}{\Lambda^4}g_8\partial_t \left( \ddot{\bar{\phi}} \dot{\bar{\phi}} \right) \ , \label{eq:HomMeff} \\
c_s^2( {\bf k}) &=1- \frac{8}{\Lambda^4} g_8 \dot{\bar{\phi}}^2 - \frac{8}{\Lambda^8}g_{12}|{\bf{k}}|^2 \ddot{\bar{\phi}}^2 +\frac{96}{\Lambda^8} g_8^2 \dot{\bar{\phi}}^4   \, , \label{eq:HomSpeed}
\end{align}
up to terms that are more suppressed.
Note that the $g_{10}$ contribution to the square sound speed vanishes as expected as the quartic Galileon\footnote{As expected, if one were to choose the parametrisation for the $g_{10}$ operator in \eqref{eq:L} where the term $\Box \phi (\p \phi)^2$ is removed by field redefinition, the $g_{10}$ contribution to $c_s^2({\bf k})$ would be a total derivative and would also vanish at the level of the time delay so long as one considers background profiles with vanishing boundary terms, as is done in our analysis.} vanishes on an effectively one-dimensional background. To analyse this term, one needs to explore backgrounds that are effectively at least two-dimensional and in the next section we will consider a spherically-symmetric background\footnote{Cylindrically-symmetric background were also considered and lead to no additional insights, we shall therefore not present them in this work.}.

Note that to stay within the regime of validity of the EFT we require that
\begin{equation}
\frac{H\bar{\Phi}_0}{\Lambda^2} \ll 1 \, , \quad  \frac{H}{\Lambda}\ll 1 \, , \quad \text{and} \quad  \frac{\omega H}{\Lambda^2} \ll 1 \ ,
\end{equation}
where $\bar{\Phi}_0$ is the overall scale of the background field, or one can take $\bar{\Phi}_0={\rm max}(|\bar\phi(t)|)$.
As a consequence, this ensures that $c_s \sim 1$, up to small perturbative corrections. Furthermore, the validity of the WKB regime where the perturbations vary much faster than the background implies that $| {\bf k}|H^{-1}\gg 1$. These requirements imply that the speed \eqref{eq:HomSpeed} is subluminal for $g_8>0, g_{12}=0$ and for $ g_{8}=0, g_{12}>0$. Even though the departure from the speed of light will be small, this effect may pile-up when dealing with large observation times and lead to macroscopic effects. To understand how this could occur, we establish the amount of support $\Delta x$ the field would be able to gain outside the standard Minkowski light-cone
\begin{equation}
\Delta x=H^{-1} \int_{\tau_i}^{\tau_f} (1-c_s(\tau)) \mathrm{d} \tau \ ,
\end{equation}
where we introduced the dimensionless time $\tau=H t$. The light-cone observed by the perturbation is smaller than the Minkowski one by $\Delta x$. Hence, violations of causality arise for waves with three-momentum $\bf{k}$  enjoying $|{\bf{k}}|\Delta x<-1$ for any $\Delta \tau=\tau_f-\tau_i>0$ while remaining within the regime of validity of the EFT.  That is, if the distance that the perturbations can propagate outside of the Minkowski light-cone becomes larger than the wavelength of the perturbation. As is well known, in this setup, there is no risk of causality violation  if $g_8>0$ and $g_{12}>0$.
However if $g_8<0$, one can easily find solutions on which $|{\bf{k}}|\Delta x<-1$.

Consider for instance a time-localised profile of the form $\bar \phi(\tau)=\bar{\Phi}_0 e^{-\tau^2}$. The resulting support outside the  light-cone will then be
\begin{equation}
|{\bf k}|\Delta x=\frac{|{\bf k}|}{H} \int_{-\infty}^{\infty} (1-c_s(\tau)) \mathrm{d} \tau=4\sqrt{\pi}\frac{|{\bf k}|}{H}\left[\frac{H \bar{\Phi}_0 }{\Lambda^2}\right]^2\(g_8+3g_{12}\frac{|{\bf k}|^2H^2}{\Lambda^4}-\frac{9g_8^2}{\sqrt{2}}\frac{H^2 \bar{\Phi}_0^2}{\Lambda^4}\) \ .
\end{equation}
In the regime of validity of the EFT, $H \bar{\Phi}_0\ll \Lambda^2$  and the terms quadratic in $g_8$ are naturally negligible. While the prefactor in square brackets should be small, this can always be compensated by a sufficiently large $|{\bf k}|H^{-1}\gg 1$ as required from the validity of the WKB approximation. For those solutions the term linear in $g_{12}$ is always subdominant as $|{\bf k}| H\sim \omega H \ll \Lambda^2$.  Hence as $g_8<0$, there are solutions within the regime of validity of the EFT for which the time advance is resolvable $|{\bf k}|\Delta x<-1$ signalling a violation of causality. This result complements that derived in \cite{deRham:2020zyh}.

On the other hand for more involved profiles, the term linear in $g_{12}$ can be sufficiently enhanced so that it dominates over the term linear in $g_{8}$ despite the  $|{\bf k}|^2H^2 \Lambda^{-4}$ suppression. As a simple proof of principle, we could consider for instance a profile of the form $\bar \phi(\tau)=\bar{\Phi}_0 \tau^{2} e^{-\tau^2}$ for which the support then becomes
\ba
|{\bf k}|\Delta x=\frac{7\sqrt{\pi}}{2\sqrt{2}}\frac{|{\bf k}|}{H} \left[\frac{H \bar{\Phi}_0 }{\Lambda^2}\right]^2\(g_8+\frac{57}{7}g_{12}\frac{|{\bf k}|^2H^2}{\Lambda^4}-\frac{6129}{1792}\frac{g_8^2}{\sqrt{2}}\frac{H^2 \bar{\Phi}_0^2}{\Lambda^4}\)\,,
\ea
hence taking $|{\bf k}| H/\Lambda^2 \sim 0.2$ ensures validity of the EFT, while the $g_{12}$ dominates over the $g_8$ term and hence a resolvable time advance will be possible within the regime of validity of the EFT for negative $g_{12}\sim -1$ even if $g_8\sim 1$. One could push the analysis to more generic profiles and derive a more systematic resolvable support outside the light-cone whenever $g_{12}$ is negative however at this stage moving on to  spherically symmetric profiles will prove more instructive and positivity of $g_{12}$ from pure causality considerations will be proven in that context (see the summary of the causality constraints depicted in Fig.~\ref{fig:bounds} where it is clear that even in the presence of a generic positive $g_8$, $g_{12}$ still ought to be positive to ensure causality on generic configurations that remain in the regime of validity of the EFT). No additional information can be obtained from this analysis since $g_{10}$ has no effects on the dispersion relation as explained under Eq.~\eqref{eq:HomSpeed} and, as argued previously, the $g_8$ and $g_{12}$ contributions to the speed are sign definite so we cannot bound these coefficients from above.

\section{Spherically-symmetric background} \label{sec:Spherical}
We proceed to explore causality constraints on a static and spherically-symmetric background $\bar{\phi}(r)$ for which the operator $g_{10}$ is relevant. This  allows us to establish to which extent the non-linear positivity bounds in Eq.~\eqref{eq:pos} can be reproduced using causality considerations at low energy without other information from its UV completion. Given the symmetries of the background, we perform an expansion in spherical harmonics\footnote{Due to the azimuthal symmetry, we can neglect the $\varphi$ dependence of the spherical harmonics and work with the Legendre polynomials.} (partial waves) and write our perturbation as $\psi=\sum_{\ell} e^{i\omega t}Y_{\ell}(\theta) \delta \rho_{\ell}(r)$. We obtain an equation of motion for the $\ell$-mode radial perturbation, $\rho_{\ell}$, which schematically is
\begin{equation}
	\delta\rho''_{\ell}(r)+A(\omega^2,r)\delta\rho'_{\ell}(r)+\left(\omega^2 C(\omega^2,r)-\frac{\ell(\ell+1)}{r^2}+B(r,\ell) \right)\delta\rho_{\ell}(r)=0 \, . \label{eq:radialeomFriction}
\end{equation}
In the absence of interactions we have $B(r,\ell)=0$ (up to the mass of the scalar field which we treat as negligible). We perform a field redefinition, $\delta\rho_{\ell}(r)=e^{-\int A(\omega^2,r)/2 \mathrm{d}r}\chi_{\ell}(r)$, to remove the friction term and get
\begin{equation}
	\chi''_{\ell}(r)+\frac{1}{c_s^2(\omega^2,r)}\left(\omega^2-V_{\text{eff}}\right)\chi_{\ell}(r)=0 \ , \quad {\rm with}\quad  V_{\text{eff}}\equiv \frac{\ell(\ell+1)}{r^2}+\tilde{B}(r,\ell) \, . \label{eq:radialeom}
\end{equation}
We can obtain this equation in an exact form, but for our purposes, we will consider an expansion in the parameters $\delta_1$ and $\delta_2$ defined in Eq.~\eqref{eq:validity}. Let us consider a spherically-symmetric background of the form $\bar\phi(r)=\bar{\Phi}_0 f(r)$ and change coordinates to $R=r/r_0$, where $r_0$ is an arbitrary length scale that measures the variation of the background and $\bar{\Phi}_0$ has dimensions of mass,
\begin{equation}
	\bar\phi(r/r_0) = \bar{\Phi}_0 f(R) \,.
	\label{eq:deff}
\end{equation}
With these definitions, both the profile $f$ and the radius $R$ are dimensionless. Note that the definitions of all dimensionless parameters and functions are reported in Table~\ref{tab:dictionaryDimensionless} of Appendix \ref{ap:DefDimLess}. Moreover, we expect $f$ and its derivatives $f^{(n)}$ (where the differentiation is taken with respect to $R$) to be at most of $\mathcal{O}(1)$. The validity of the EFT implies that:
\begin{equation}
	\epsilon_1\equiv\frac{\bar{\Phi}_0}{r_0\Lambda^2}\ll1 \, , \quad  \epsilon_2\equiv\frac{1}{r_0 \Lambda}\ll1 \, , \quad  \text{and} \quad \Omega \epsilon_2\equiv \frac{\omega}{r_0\Lambda^2} \ll 1 \ . \label{eq:eps}
\end{equation}
At the level of the phase shift, each contribution from the $g_i$ terms would scale as follows
\begin{equation}
	g_8 : \mathcal{O}\left( \epsilon_1^2\right), \qquad g_{10} : \mathcal{O} \left(\epsilon_1^2 \epsilon_2^2\right), \qquad g_{12} : \mathcal{O} \left(\epsilon_1^2\epsilon_2^2\Omega^2\right) \,.
\end{equation}
More generally, any term coming from $g_8^{n_1} g_{10}^{n_2} g_{12}^{n_3}$ will be suppressed by at least a power
\begin{equation*}
	\epsilon_1^{2(n_1+n_2+n_3)} \epsilon_2^{2(n_2+n_3)}\Omega^{2n_3} \ .
\end{equation*}
We write our expressions in terms of $\epsilon_1$, $\epsilon_2$, and $\Omega$ and then perform an expansion up to order $\epsilon_1^2\epsilon_2^2$ and  $\epsilon_1^4$, which requires the assumptions $\epsilon_1^2\ll\epsilon_2$ and $\epsilon_2^2\ll\epsilon_1$. In fact, we will consider $\epsilon_1$ and $\epsilon_2$ of the same order, but allow the freedom of the exact value of these scales to be different. We will refer to these contributions as the leading order (LO) or $\mathcal{O}(\epsilon^4)$. Thus, we need to keep track of the contributions coming from the $g_8, \ g_{10}, \ g_8^2, \ g_{12}$ terms. Expanding up to order $\epsilon_1^2\epsilon_2^2$ we find
\begin{equation}
	\chi''_{\ell}(R)+W_{\ell}\chi_{\ell}(R)=0 \, , \quad W_{\ell}=\frac{(\omega r_0)^2}{c_s^2(\omega^2,R)}\left(1-\frac{V_{\text{eff}}(R)}{(\omega r_0)^2}\right) \, ,  \label{eq:WKBR}
\end{equation}
where prime now denotes a derivative with respect to $R$. Note that the expansion is not just in $\epsilon$, but in $g_i\epsilon$. For this perturbative result to be correct, we need to make sure higher-order corrections to the series expansion in Eq.~\eqref{eq:WKBR} above are small. Thus, we require schematically that $g_i \epsilon \ll 1$, which as expected, simply tells us that we should not consider too large values for the Wilson couplings. Wilson coefficients much larger than unity should be rescaled appropriately in the cutoff $\Lambda$ resulting in a lower cutoff scale. We now  solve Eq.~\eqref{eq:radialeom} using the WKB approximation, first analysing how far in the WKB approximation one needs to include contributions to be consistent with the EFT expansion. Once the consistency of the WKB expansion with the EFT expansion is established, we can then explore the parameter space in which causality violations can arise. We will do so for the cases $\ell=0$ and $\ell\neq 0 $ separately.

\subsection{Regime of validity of the WKB approximation}
We start by considering Eq.~\eqref{eq:WKBR} given in terms of dimensionless variables
\begin{equation}
	\chi_{\ell}''(R) + (\omega r_0)^2 \hat{W}_{\ell}(R) \chi_{\ell}(R) = 0 \,, \qquad \hat{W}_{\ell}(R) = \frac{W_{\ell}(R)}{(\omega r_0)^2} = \frac{1}{c_s^2(\omega^2,R)}\left(1-\frac{V_{\text{eff}}(R)}{(\omega r_0)^2}\right) \,. \label{eq:eomDim}
\end{equation}
Since we assume that the perturbation fluctuates faster than the background, namely,
\begin{equation}
\frac{	\lambda_{\text{perturbation}} }{\lambda_{\text{background}}}=\frac{1}{\omega r_0} =\frac{\epsilon_2}{\Omega}\ll 1 \ , \label{eq:WKBreq}
\end{equation}
we can solve the equation above using the WKB method. In this approach, the solution to the equation of motion up to $n$th-order correction in the WKB formula is given by
\begin{equation}
	\chi_{\ell}^{(n)}(R) \propto \left( e^{i (\omega r_0) \int_0^R \sum_{j \geq 0}^n \delta_{\rm WKB}^{(j)} \mathrm{d}R} - e^{- i (\omega r_0) \int_0^R \sum_{j \geq 0}^n \delta_{\rm WKB}^{(j)} \mathrm{d}R} \right) \,,
\end{equation}
where the boundary conditions were chosen such that $\chi_{\ell}^{(n)}(R=0)=0$ and  $\delta_{\rm WKB}^{(j)}$ is the $j$th-order term in the WKB series expansion whose explicit expressions can be found in \cite{WKBbook} and we list the relevant ones for our analysis below. Noting that $\hat{W}_{\ell}>0$, it is easy to realise that $\delta_{\rm WKB}^{(j)}$ are purely imaginary total derivatives when $j$ is odd, meaning that they do not contribute to the phase but simply to the amplitude. In the end, we have that the phase is proportional to
\begin{equation}
	\sum_{j \geq 0} \delta_{\rm WKB}^{(2j)} = \delta_{\rm WKB}^{(0)} + \delta_{\rm WKB}^{(2)} + \delta_{\rm WKB}^{(4)} + \cdots \,,
\end{equation}
where the first three contributions are
\begin{align}
	\delta_{\rm WKB}^{(0)} &= \sqrt{\hat{W}_{\ell}} \,, \\
	\delta_{\rm WKB}^{(2)} &= - \frac{1}{(\omega r_0)^2} \frac{1}{8\sqrt{\hat{W}_{\ell}}} \left( \frac{\hat{W}''_{\ell}}{\hat{W}_{\ell}} - \frac{5}{4} \left( \frac{\hat{W}'_{\ell}}{\hat{W}_{\ell}} \right)^2 \right) \,, \\
	\delta_{\rm WKB}^{(4)} &= \frac{1}{(\omega r_0)^4} \frac{1}{32 \hat{W}_{\ell}^{3/2}} \left[ \frac{\hat{W}^{(4)}_{\ell}}{\hat{W}_{\ell}} - 7 \frac{\hat{W}'_{\ell} \hat{W}^{(3)}_{\ell}}{\hat{W}^2_{\ell}} - \frac{19}{4} \left( \frac{\hat{W}''_{\ell}}{\hat{W}_{\ell}} \right)^2 + \frac{221}{8} \frac{\hat{W}''_{\ell} \hat{W}'_{\ell}{}^2}{\hat{W}^3_{\ell}} - \frac{1105}{64} \left( \frac{\hat{W}'_{\ell}}{\hat{W}_{\ell}} \right)^4 \right] \,.
	\label{eq:WKBcorrec}
\end{align}
For our purposes, we need to consider terms up to order $\mathcal{O}(\epsilon^4)$. Below, we will see that $\delta_{\rm WKB}^{(2)}$ has contributions of order $\epsilon_1^2/(\omega r_0)^2=\epsilon_1^2\epsilon_2^2/\Omega^2$ that should be taken into account in order to have a consistent expansion at leading-order (LO) for the phase shift, and hence the time delay. In some cases, we would be interested in computing the next EFT contribution to ensure that it is a small effect that does not change our bounds. These next-to-leading (NLO) corrections include terms of the following orders $\mathcal{O}(\epsilon_1^6, \epsilon_1^2 \epsilon_2^4, \epsilon_1^4 \epsilon_2^2)$. In this case, we will need to include $\delta_{\rm WKB}^{(4)}$ corrections to the WKB formula.

We can establish the validity of the WKB approximation by looking at the relative error between the exact solution $\chi_{\ell}$ and the WKB approximation up to $n$th-order corrections $\chi_{\ell}^{(n)}$. Thus we require that
\begin{equation}
\frac{\chi_{\ell}(R)-\chi_{\ell}^{(n)}(R)}{\chi_{\ell}(R)}\sim\frac{1}{(\omega r_0)^n}\int_0^R \delta_{\rm WKB}^{(n+1)} \mathrm{d}R \ll 1 \,,
\end{equation}
as well as
\begin{equation}
\frac{1}{(\omega r_0)^n}\int_0^R \delta_{\rm WKB}^{(n+1)} \mathrm{d}R	\ll \frac{1}{(\omega r_0)^{n-1}}\int_0^R \delta_{\rm WKB}^{(n)} \mathrm{d}R \,,
\end{equation}
in order for the WKB to be a useful approximation given by an asymptotic series in $(\omega r_0)^{-1}$ \cite{WKBbook}. Similarly, we want to ensure that the next order WKB terms are indeed negligible at the order in the perturbative expansion that we are working on. This can be checked by computing the next order in Eq.~\eqref{eq:eomDim} inferring that,
\begin{equation}
{\chi^{(n)}_{\ell}}''(R) + (\omega r_0)^2 \hat{W}_{\ell}(R) \chi^{(n)}_{\ell}(R) = 	\mathcal{E}_{\ell}^{(n+1)}\sim  \frac{\delta_{\rm WKB}^{(n+1)}}{\delta_{\rm WKB}^{(0)}} \ .
\end{equation}
From which we can see that the leftover is of order $ \mathcal{O}((\omega r_0)^{-(n+1)})$ which is small provided $(\omega r_0) \gg 1$, as postulated earlier. In practice, we compute carefully the order of these leftover and make sure that it vanishes at LO.

\subsection{Case 1: Monopole} \label{sec:monopole}
In this section, we analyse the causality bounds on the EFT Wilson coefficients that arise when scattering the monopole mode. To do so, we consider Eq.~\eqref{eq:WKBR} with $\ell=0$. At leading order, the function $\hat{W}_0$ that appears in the equation of motion reads
\begin{align}
	\left. \hat{W}_0(R) \right|_{\rm LO} =& 1+8 g_8 \epsilon _1^2 f'(R)^2+96 g_8^2 \epsilon _1^4 f'(R)^4+8 g_{12} \Omega ^2 \epsilon _1^2 \epsilon_2^2 f''(R)^2 +24 g_{10} \epsilon _1^2 \epsilon _2^2 \frac{f'(R) f''(R)}{R}\nonumber \\
	&+12 \frac{g_8}{\Omega ^2} \epsilon _1^2 \epsilon _2^2 \left(2 \frac{f'(R) f''(R)}{R}+\frac 12 \p_R^2 f'(R)^2\right) \,,
\end{align}
where $f$ and $R$ are respectively the dimensionless spherically-symmetric background and radius defined in \eqref{eq:deff}. As explained earlier we have performed an expansion in the small dimensionless parameters $\epsilon_1$, $\epsilon_2$, $\Omega\epsilon_2$ that measure the validity of the EFT. There is another dimensionless parameter which measures the validity of the WKB expansion and can be written in terms of the previous parameters, namely, $(\omega r_0)^{-1}=\epsilon_2/\Omega$. In order to obtain tight bounds for the Wilsonian coefficients one needs to consider the extreme situation where these small parameters are as large as possible while maintaining the EFT under control and being able to compute the necessary WKB corrections at this order. We will be computing the time delay at LO while ensuring validity of the EFT  by imposing Eq.~\eqref{eq:eps}. These requirements together with the validity of the WKB approximation lead to $\epsilon_2\ll\Omega \ll 1/ \epsilon_2$, but in practice, we require slightly tighter bounds given by $\sqrt{\epsilon_2}<\Omega < 1/\sqrt{ \epsilon_2}$,
together with $\epsilon_1^2<\epsilon_2$ and $\epsilon_2^2<\epsilon_1$ in order to have a well-defined expansion truncated at $\mathcal{O}(\epsilon^4)$. For example, this means that we keep corrections of order $(\epsilon_1/(\omega r_0))^2$ but neglect $(\epsilon_1/(\omega r_0)^2)^2$. The latter type of corrections arise in the effective potential, but not in the speed.\\

It is instructive to look at the sound speed and effective potential which, at leading order, are given  by
\begin{align}
	&\left. c_s^2(\omega^2,R) \right|_{\rm LO}=1-8 g_8 \epsilon _1^2 f'(R)^2-32 g_8^2 \epsilon _1^4 f'(R)^4 - 8 g_{12} \epsilon _1^2 \epsilon _2^2 \frac{\omega^2}{\Lambda ^2} f''(R)^2 -24 g_{10} \epsilon _1^2 \epsilon _2^2 \frac{f'(R) f''(R)}{R}  \, ,\nn \\
	&\left. V_{\text{eff}}(R) \right|_{\rm LO}= -12 g_8 \epsilon _1^2 \left( 2 \frac{f'(R)f''(R)}{R} + \frac 12 \p_R^2 f'(R)^2  \right)   \, .
\end{align}
One should note here that the effective potential term is suppressed by $(\omega r_0)^{-2}=\epsilon_2^2 \Omega^{-2}$ with respect to the sound speed term. The corrections at NLO to the speed of sound and to the effective potential are listed in Appendix \ref{ap:NLO}. From the sound speed expression, we can understand whether we should expect to be able to reproduce any of the positivity bounds in Eq.~\eqref{eq:pos}. Firstly, as already argued in Section \ref{sec:speed}, the $g_8$ and $g_{12}$ contributions to the speed are clearly sign definite. Next, we analyse the $g_{10}$ contribution. This term appears to be sign indefinite, but under the integral, it is equivalent to a sign definite contribution up to total derivatives that will vanish at the boundaries. Hence, with use of the monopole, we can only expect to be able to bound the $g_{10}$ coefficient from below and we will need to resort to higher multipoles to bound $g_{10}$ from above. \\

We can now determine the phase shift experienced by the perturbation travelling in the spherically-symmetric background. For that we first rewrite the solution to the perturbed equation of motion in the following way
\begin{equation}
	\chi^{(n)}_0(R) \propto e^{-i (\omega r_0) \int_0^R \left( \sum_{j \geq 0}^n \delta_{\rm WKB}^{(j)} - 1 \right) \mathrm{d}R} \left( e^{2 i (\omega r_0) \int_0^R \left( \sum_{j \geq 0}^n \delta_{\rm WKB}^{(j)} -1 \right) \mathrm{d}R} e^{i (\omega r_0) R} - e^{-i (\omega r_0) R} \right) \,,
\end{equation}
so that it can be compared to the asymptotic solution $\chi_0(R) \propto \left( e^{2i \delta_0} e^{i (\omega r_0) R} - e^{-i (\omega r_0) R} \right)$
to find that the expression for the phase shift at $\ell=0$ reads
\begin{equation}
	\delta_0(\omega) = \omega r_0\int_0^{\infty} \left( \sum_{j \geq 0} \delta_{\rm WKB}^{(j)}- 1 \right) \mathrm{d}R  \,, \label{eq:phaseshift}
\end{equation}
which is positive for $0 < c_s < 1$ and large enough $\omega r_0$ as seen when using Eqs.~\eqref{eq:eomDim} and \eqref{eq:WKBcorrec}:
\begin{equation}
	\delta_0(\omega)\sim \omega r_0\int_0^{\infty} \left( \frac{1}{c_s}- 1 \right) \mathrm{d}R  \, .
\end{equation}
From Eq.~\eqref{eq:phaseshift}, we see that the dimensionless time delay of a partial wave with zero angular momentum is given by
\begin{equation}
	\omega \Delta T_0(\omega) = 2 \omega \frac{\p\delta_0(\omega)}{\p \omega} = 2\omega \int_0^{\infty}\frac{\p}{\p \omega}\left( (\omega r_0) \left( \sum_{j \geq 0} \delta_{\rm WKB}^{(j)} - 1 \right)\right)  \mathrm{d}R \equiv  \int_0^{\infty}\mathcal{I}_0(\omega,R) \mathrm{d}R \, ,
\end{equation}
where up to $\mathcal{O}(\epsilon^4)$ we have
\begin{align}
	\left. \mathcal{I}_0(\omega^2,R) \right|_{\rm LO}= &8 (\omega r_0) \epsilon_1^2 \left[ g_8 f'(R)^2 + 10 g_8^2 \epsilon_1^2 f'(R)^4 +3 g_{12} \Omega^2 \epsilon_2^2 f''(R)^2 \vphantom{\frac{1}{2}} \right. \nonumber \\
	& \left. - \frac{g_8}{\Omega^2} \epsilon_2^2 \left( 3 \frac{f'(R)f''(R)}{R} + \frac 12 \p_R^2(f'(R)^2) \right) + 3 g_{10} \epsilon_2^2 \frac{f'(R) f''(R)}{R} \right] \, .
\end{align}
As appropriate for a scattering regime which is intrinsically wave-like such as the $\ell=0$ case,  we are computing the time delay at fixed $\ell$. As mentioned earlier, we will consider background profiles giving null boundary terms so that we can neglect any contribution from total derivative terms. Taking this into consideration and performing integration by parts we find that the above equation can be written as
\begin{align}
	\left. \mathcal{I}_0(\omega^2,R) \right|_{\rm LO}=&\ 8 (\omega r_0) \epsilon_1^2 \left[ g_8 f'(R)^2 + 10 g_8^2 \epsilon_1^2 f'(R)^4 + 3 \epsilon_2^2 \left( g_{12} \Omega^2 f''(R)^2 + \frac{1}{2} \left( g_{10} - \frac{g_8}{\Omega^2} \right) \frac{f'(R)^2}{R^2} \right) \right] \nonumber \\
	&+ \text{total derivatives}\, . \label{eq:TDintegrand}
\end{align}
We can now explicitly see that the contribution from each term in the EFT expansion is sign definite when looking at the scattering of $\ell=0$ modes. From these expressions and the constraints from the validity of the EFT and the WKB approximation in Eqs.~\eqref{eq:eps},~\eqref{eq:WKBreq}, we can easily see that the $g_8$ and $g_{12}$ terms can give rise to resolvable time delays. In fact, the time delay is positive for $g_8>0$ when $g_{10}=g_{12}=0$ and for $g_{12}>0$ when $ g_{8}=g_{10}=0$, however we will soon be able to make more general statements. For the $g_{10}$ terms, one can also obtain a resolvable time delay, but this requires tuning of the function $f$ to make the time delay large while satisfying Eq.~\eqref{eq:eps}. After considering the high $\ell$ case in the following section, we will analyse the situations when a resolvable time advance can occur in section~\ref{sec:theory}.

\subsection{Case 2: Higher-order multipoles} \label{sec:multipole}
We will now consider the case of partial waves with $\ell>0$. As first noted by Langer \cite{PhysRev.51.669}, the standard WKB approach fails to be useful when considering low multipole contributions since the approximation fails to reproduce the behavior of the solutions near $r=0$. To deal with this, one can perform a change of variable, $r=e^{\rho}$, in order to map the singularity $r=0$ to $\rho=-\infty$. Then, the exponentially decaying WKB solution reproduces the correct asymptotics at $\rho=-\infty$. We proceed to change the variables in Eq.~\eqref{eq:radialeomFriction} as described above and obtain an equation of motion that contains a friction term which we remove with a field redefinition to get,
\begin{equation}
	\p_{\rho}^2 \delta \rho_{\ell}(\rho)= - \widehat{W}_{\ell}(\rho) \delta \rho_{\ell}(\rho)\,.
\end{equation}
Then, we solve this equation using the WKB approximation. To find the phase shift, we want to express $\widehat{W}_{\ell}(\rho)$ back in terms of the dimensionless radial coordinate $R=r/r_0$. For generic multipole we define the dimensionless quantity
\begin{equation}
W_{\ell}(R)\ \equiv \frac{1}{(\omega r)^2} \widehat{W}_{\ell}(\rho(r)) \,,
	\label{eq:defWlr}
\end{equation}
 note that this is not precisely the same definition as what was performed in \eqref{eq:eomDim}. Here
 the factor $1/r^2$ captures the Jacobian of the transformation:
  \be
  \int \sqrt{\widehat{W}_{\ell}(\rho)} \mathrm{d}\rho  = \omega r_0 \int \sqrt{W_{\ell}(r)} \mathrm{d}R \, .
  \ee

Before moving on, we note that within the present formalism we cannot compute the time delay beyond the leading-order WKB approximation for the $\ell>0$ case. It is well known that higher-order ($n>0$) WKB corrections are divergent at the turning point. This simply signals the breaking of the approximation in this region and the WKB solution can be improved by matching to an asymptotic solution near the turning point. Nevertheless, this does not modify the asymptotic behavior of the WKB solution and thus does not change the inferred time delay. While these subleading contributions seem to involve infinities at finite order in the WKB series expansion, the physical phase shift is finite so that upon appropriate re-organization or  resummation of the series the result will end up being finite. We could in principle carry out this resummation or re-organization of the series, however  for simplicity  we focus here instead in the regime where $(\omega r_0) \gg 1$ so that all WKB corrections can safely be ignored.\\

At first order in the WKB approximation, the phase shift of the partial wave with $\ell>0$ can be identified as (see Appendix B of \cite{deRham:2020zyh})
\begin{equation}
	\delta_{\ell} = (\omega r_0) \left[ \int_{R_t}^{\infty} \mathrm{d}R \left( \sqrt{W_{\ell}(R)} - 1 \right) - R_t + \frac{1}{2} B \pi \right] \, , \label{eq:PSl}
\end{equation}
where the  $R_t=r_t/r_0$ is the dimensionless turning point such that $W_{\ell}(r_t)=0$ and we have introduced the dimensionless impact parameter $B = b/r_0$, where $b = (\ell + 1/2)/ \omega$ is the impact parameter of the free theory, \ie when $g_{i}=0$. We remind the reader that the definitions of all dimensionless parameters are reported in Table \ref{tab:dictionaryDimensionless} of Appendix~\ref{ap:DefDimLess}.\\

In order to perform the integral in Eq.~\eqref{eq:PSl} analytically, we will expand the integrand at LO as in the monopole case. We have to be careful when splitting the integral order by order so that each term is a converging integral. To do so, we start by writing
\begin{equation}
	W_{\ell}(R) =	W_{\ell}(R)|_{{g_i=0}}+\delta W_{\ell}(R)\ ,
\end{equation}
where $	W_{\ell}(R)|_{{g_i=0}}$ is the contribution arising purely from the angular momentum contributions but no self-interactions. Using the fact that $W_{\ell}(R_t)=0$, this can be rewritten as
\begin{equation}
	W_{\ell}(R) =\left(1 - \frac{R_t^2}{R^2}\right) + \Delta W_{\ell}(R)\ , \quad   \Delta W_{\ell}(R)=\delta W_{\ell}(R)-\frac{R_t^2}{R^2}\delta W_{\ell}(R_t)  \ ,
\end{equation}
so that each contribution is finite at the integration boundaries and the integrals at each order in $\epsilon_1$ and $\epsilon_2$ converge. Now, expanding the square root at $\mathcal{O}(\epsilon^4)$ gives
\begin{equation}
	\sqrt{W_{\ell}(R)} = \sqrt{ 1 - \frac{R_t^2}{R^2} }  +  \frac{U_{\ell}(R)}{\sqrt{ 1 - \frac{R_t^2}{R^2} }} \,,
\end{equation}
where $U_{\ell}(R_t)=0$. The LO explicit expressions for $W_{\ell}(R)$, $U_{\ell}(R)$, and $R_t$ can be found in Appendix \ref{ap:NLOmultipole}. Integrating this expression gives the phase shift at $\mathcal{O}(\epsilon^4)$. Note that $R_t = B + \mathcal{O}(\epsilon^4)$ and $U_{\ell} = \mathcal{O}(\epsilon^2)$, hence, when dealing with the $U_{\ell}$ term, the turning point $R_t$ can be replaced by $B$ since any corrections will contribute at NLO. This means that we can write
\begin{equation}
	\int_{R_t}^{\infty} \left( \sqrt{W_{\ell}(R)} -1 \right) \, \mathrm{d}R = \int_{R_t}^{\infty} \left( \sqrt{ 1 - \frac{R_t^2}{R^2} } -1 \right) \, \mathrm{d}R + \int_B^{\infty} \frac{U_{\ell}(R)}{\sqrt{ 1 - \frac{B^2}{R^2} }} \, \mathrm{d}R \,,
\end{equation}
giving
\begin{equation}
	\delta_{\ell}(\omega) = (\omega r_0) \left[ \int_B^{\infty} \frac{U_{\ell}(R)}{\sqrt{ 1 - \frac{B^2}{R^2} }} \, \mathrm{d}R + \frac{\pi}{2} \left( B - R_t \right) \right] \,. \label{eq:PSlargeL}
\end{equation}
To get the time delay, we need to differentiate the expression above with respect to $\omega$. As opposed to the monopole case where we fixed $\ell$, when going to higher multipoles it is convenenient to think of the scattering not in terms of the scattering of waves but of particles specified by a given impact parameter. That is to say, what is naturally held fixed for particle scattering is the impact parameter $b$ (or $B$). This is the time delay traditionally considered in the eikonal approximation (see for example \cite{Camanho:2014apa}). Thus, the time delay reads
\begin{equation}
	(\omega \Delta T_{b}(\omega)) = 2 \frac{\partial \delta_{\ell}(\omega)}{\partial \omega} \big|_{b}= 2 (\omega r_0) \left[ \int_{R_t}^{\infty} \left( \p_{\omega} \left( \omega \sqrt{W_{\ell}(R)} \right) - 1 \right) \mathrm{d}R - R_t + \frac12 B \pi \right] \, ,
\end{equation}
which after using Eq.~\eqref{eq:PSlargeL} can be written as
\begin{equation}
	(\omega \Delta T_{b}(\omega)) = 2 (\omega r_0) \left[ \int_{B}^{\infty} \frac{\p_{\omega} (\omega U_{\ell}(R))}{\sqrt{1 - \frac{B^2}{R^2}}} \mathrm{d}R + \frac{\pi}{2} \left( B - \p_{\omega} (\omega R_t) \right) \right] \,. \label{eq:TDlargel}
\end{equation}

In the next section, we will explore the regions in Wilson coefficient space that can lead to a resolvable time advance given by Eq.~\eqref{eq:TDlargel}. Contrary to the $\ell=0$ case, the contribution to the time delay from $g_{10}$, found in Eq.~\eqref{eq:velocityHighL}, is not sign definite when we have angular momentum. This will allow us to bound the $g_{10}$ coefficient from above and below. The tightest bounds will arise from considering the scattering of higher-order multipole modes. Note that while we can take the large-$\ell$ limit, we cannot take $\ell\rightarrow\infty$. This can be seen by writing $L=\ell+1/2$, and
\begin{equation}
	L =\omega b=\frac{B \Omega}{\epsilon_2} \,.
\end{equation}
The impact parameter $B$ cannot be taken to infinity, otherwise there would be no scattering. Meanwhile, $\Omega$ is bounded by Eq.~\eqref{eq:eps} so that we stay within the regime of validity of the EFT. Thus, at a fixed impact parameter, the angular momentum has an upper bounded given by
\begin{equation}
L\ll \frac{B}{\epsilon_2^2} \ .
\end{equation}
Note that this large angular momentum limit is related to the standard approach of computing phase shifts by looking at the eikonal limit of $2 \rightarrow 2$ scatterings.
\subsection{Causal shift-symmetric theories}
\label{sec:theory}
We now consider a specific background profile to obtain constraints on the Wilson coefficients given by causality. We use an analytic function in order to avoid any possible divergences at $R=0$. Furthermore, we require that the background vanishes at infinity to have a well-defined scattering around an asymptotically flat background, that is, the light-cones observed by the perturbation approach the Minkowski ones near infinity. Therefore, we will consider a profile of the form
\begin{equation}
	f(R^2)=\left(\sum_{n=0}^{p} a_{2n} R^{2n} \right) e^{-R^2} \ ,
	\label{eq:profilefR2}
\end{equation}
where $a_{2n}$ are arbitrary coefficients of order $1$. For a generic scalar field EFT in its own right (not coupled to gravity), one can always consider an external  source $J$ that would generate such a  profile. In more specific contexts where the scalar field is considered to be diagnosing one of the degrees of freedom of gravity (as would for instance be the case for the helicity-0 mode in \cite{Dvali:2000hr} or in massive gravity \cite{deRham:2010ik,deRham:2010kj}), one may consider more carefully how such a profile could be generated as discussed in Appendix~\ref{app:Matter_coupling}.\\

When considering such profiles \eqref{eq:profilefR2}, the largest contributions to the time delay (or advance) come from small powers $n$, so in practice we truncate the series by choosing $p=3$, \ie including terms up to $a_6$. Given this profile, we can explore the regions where one can obtain a resolvable time advance, that is, $\omega \Delta T_{b} \lesssim -1$ while maintaining the EFT under control and hence violate causality.
Since we have already established in the homogeneous case that $g_8$ ought to be positive we can set $g_8=1$ without loss of generality (this simply corresponds to a rescaling of all the Wilson coefficients by $g_8$). The case $g_8=0$ will be considered separately in what follows. We  use an extremisation procedure to find the largest region where causality is violated following the method is explained in Appendix~\ref{ap:procedure}.

\begin{figure}[!h]
	\begin{center}
		\includegraphics[width=0.5 \textwidth]{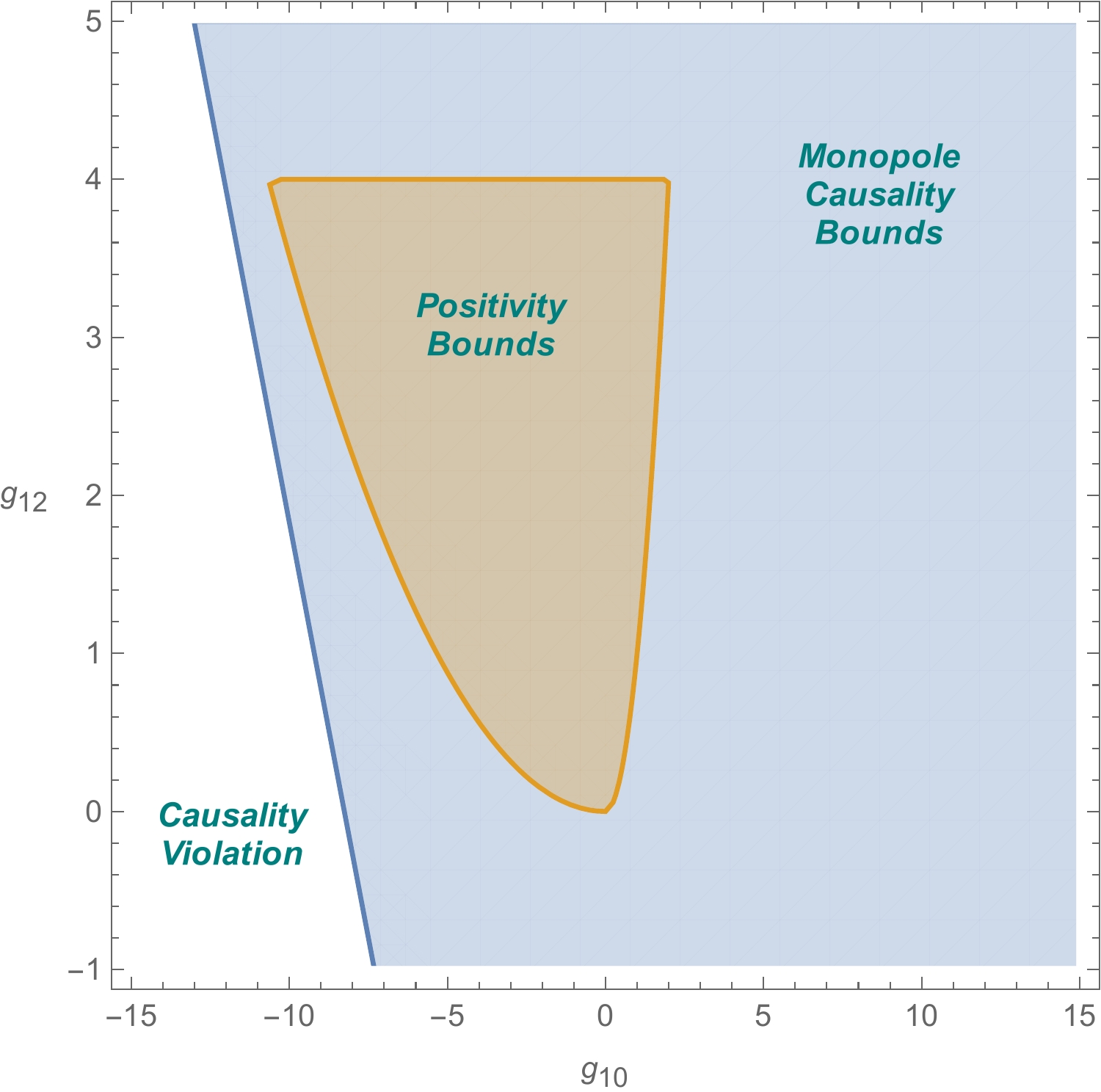}
	\end{center}
	\caption{Positivity and monopole causality constraints for the shift-symmetric scalar EFT considered in \eqref{eq:L}. In white, we observe a region that can lead to violations of causal propagation in the infrared, \ie where $\Delta T_0 < -1/\omega$. The blue region is its complement where there is no yet any indication of causality violation. Here, we have focused on bounds arising from monopole modes with a background profile given by Eq.~\eqref{eq:profilefR2} and the coefficient of the $(\p \phi)^4$ operator set to $g_8=1$. In orange, we observe the region that satisfies the positivity constraints in \cite{Tolley:2020gtv}, and assumes physical properties of the UV completion.}
	\label{fig:boundl0}
\end{figure}

\paragraph{Monopole modes:}
We first consider the $\ell=0$ case. In the extremisation procedure, we include all the constraints on the dimensionless parameters arising from the validity of the EFT as in \eqref{eq:eps} and we require that the LO and NLO results differ only by a $3\%$ for $g_{14}$ of order $1$. The NLO contributions, found in Appendix \ref{ap:NLO},  include WKB corrections as well as higher-order EFT terms. Within our parameterisation, we find the tightest constraints by considering
\begin{equation}
	a_0=1 \ ,\ a_2=0.72 \ , \ a_4\sim 0 \ , \ a_6=0.14 \ , \ \epsilon_1=0.36 \ ,\ \epsilon_2=0.35 \ , \ \Omega=0.70 \ , \label{eq:param}
\end{equation}
although it is likely that even tighter constraints could be derived if one considered other classes of profiles, the bounds we obtain here already serve as proof of principle.
The bounds arising from the previous choice are shown in blue in Fig.~\ref{fig:boundl0} together with the orange positivity bound from \cite{Tolley:2020gtv}. It is easy to prove, by examining Eq.~\eqref{eq:TDintegrand}, that the slope of the line delimitating the causal region from the acausal one is negative for any choice of coefficients when considering $\ell=0$. This means that when considering the monopole, we can only get a left-sided bound as argued earlier.

Note that our choice in Eq.~\eqref{eq:param} implies $\omega r_0\sim 2$ which does not suppress higher-order WKB corrections. Nevertheless, when working at $\mathcal{O}(\epsilon^4)$ we can safely consider this case since all the corrections $\delta_{\rm WKB}^{(2n)}$ for $n\geq 2$ will correspond to total derivatives that do not contribute to the phase shift. One can see that this is the case by looking at Eq.~\eqref{eq:WKBcorrec} and noting that after expanding in $\epsilon_1$ and $\epsilon_2$ up to order $\mathcal{O}(\epsilon^4)$ the WKB corrections will arise from $W^{(2n)}$.

\paragraph{Higher multipole modes:}
Moving to the higher multipoles, $\ell>0$, we consider the same profile as in Eq.~\eqref{eq:profilefR2}. By allowing finite values of $\ell$, the equation for the line $(\omega \Delta T_{b}) = -1$ separating regions of ``causality-violation"  has a new free parameter and now allows for a positive slope. This opens the possibility to constrain the causality region from both sides and below. We do not get a better lower-sided bound on $g_{10}$ but we do get an upper bound by considering the union of the constraints arising from a set of parameters as explained in Appendix~\ref{ap:procedure}. Remarkably, this method also sets a lower bound on $g_{12}$ which ought to be positive, in complete agreements with positivity bounds. In itself this is a remarkable statement as in the presence of the $g_8$ operator the speed is typically dominated by that term and little would be inferred from $g_{12}$.

\begin{figure}[!h]
	\begin{center}
		\includegraphics[width=0.5 \textwidth]{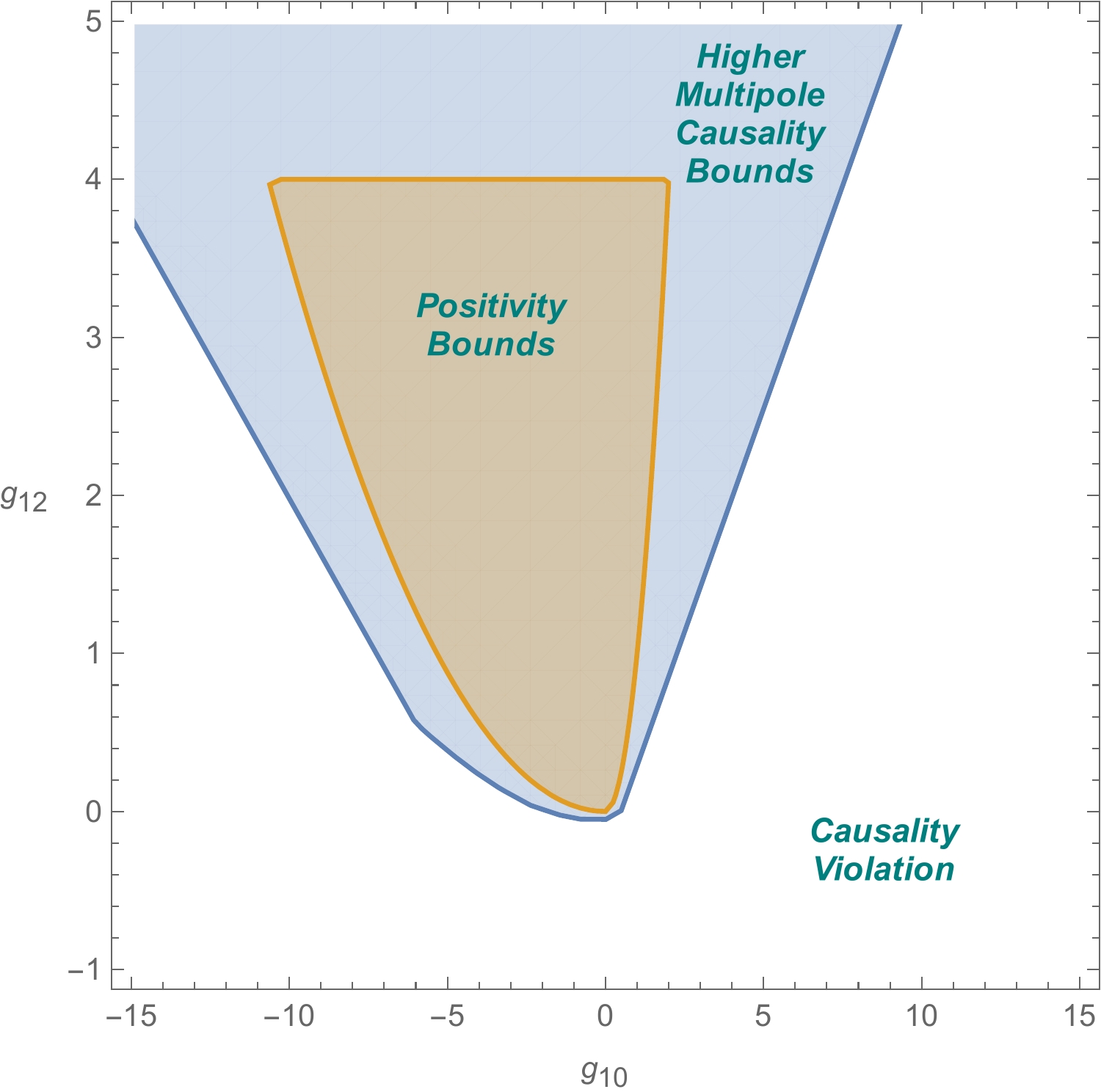}
	\end{center}
	\caption{The blue and orange regions represent the EFTs satisfying causality bounds from higher multipoles and positivity bounds respectively. The regions are computed as in Fig.~\ref{fig:boundl0} with $g_8=1$, but the causality constraints are those arising from higher-order multipole modes.}
	\label{fig:boundg81}
\end{figure}

Our results are shown in Fig.~\ref{fig:boundg81} where the causality bound region corresponds to the intersection of regions in Wilson coefficient space that do not give rise to resolvable time advances as defined in Appendix~\ref{ap:procedure}. An example of a set of parameters that we use to obtain the causality bounds is given by
\begin{equation}
	a_0=-5 \ ,\ a_2=-5 \ , \ a_4=5 \ , \ a_6=-0.91 \ , \ \epsilon_1=0.17 \ ,\ \epsilon_2=0.17 \ , \ \Omega=3 \ , \label{eq:paramLargel}
\end{equation}
leading to our tightest bound on $g_{10}$ at $g_{12}=0$. Once again we do not preclude the possibility that stronger bounds could be obtained by improved optimization methods or by considering more generic classes of profiles, however great care should be taken so as to ensure validity of the EFT and WKB approximation.  As in the previous case, we ensure that we are within the regime of validity of the EFT by satisfying Eq.~\eqref{eq:eps}. Furthermore, we only work at leading order in the WKB approximation and guarantee that higher-order corrections are negligible by taking $\omega r_0\sim\mathcal{O}(20)$. Note that, as explained earlier, odd higher-order WKB corrections only contribute to the overall amplitude and hence, corrections to the phase shift (and time delay) only come from even higher-order WKB corrections, which are then suppressed by powers of $(\omega r_0)^2\sim\mathcal{O}(400)$. In contrast to the $\ell=0$ analysis, we cannot compare to the NLO corrections since these would include WKB corrections that we cannot compute within our formalism as explained in the previous section. However, we do ensure smallness of the corrections by relying on dimension analysis given by Eq.~\eqref{eq:WKBreq}. It is interesting to note that the tightest bounds that we found come from the region where $\ell\sim\mathcal{O}(30)$, and thus are related to calculations in the eikonal limit. On the other hand, the results from the previous case ($\ell=0$) arise in the opposite regime that is less explored in the literature.\\

Combining monopole and higher multipoles causality bounds gives rise to the left panel of Fig.~\ref{fig:bounds} strongly constraining the viable region of the $\{g_{10}, g_{12}\}$ parameter space.
We highlight that there is room for our procedure to be further tightened (for instance by considering more generic backgrounds and more freedom in their parameterizations and their scaling). As a result the white regions ruled out in  Figs.~\ref{fig:boundl0} and \ref{fig:boundg81} are very likely not the most optimal bounds that one can obtain from causality but already provide a close contact with standard positivity bounds and new compact positivity bounds.

\subsection{Causality in Galileon theories}
Besides the shift-symmetric theory considered throughout this paper, one can impose a more constraining, spacetime-dependent, shift symmetry given by
\begin{equation}
	\phi\rightarrow \phi +c + b_\mu x^\mu \ ,
\end{equation}
where $c$ is a constant and $b_\mu$ a constant vector. This is the Galileon symmetry \cite{Nicolis:2008in} which arises in various contexts such as massive gravity theories, brane-world models, accelerating universes, inflationary models and alternatives to inflation \cite{deRham:2012az}. Imposing this new symmetry requires that we set $g_8=0$ in Eq.~\eqref{eq:lag}. Note that $\textit{any}$ scalar low energy EFT that enjoys a  Galileon symmetry (with no other light degrees of freedom) is forbidden by positivity bounds. Setting $g_8=0$ the positivity bounds~\eqref{eq:pos} then impose $g_{10}=g_{12}=0$. This means that when viewed as a low energy scalar EFT, a Galileon  cannot have a Wilsonian UV completion that is local, unitary, causal, and Poincar\'e invariant. Here, we would like to understand whether we can obtain similar stringent bounds from infrared causality alone with no further input on the UV completion.

The analysis follows in a similar way as the shift-symmetric case above, with the only modification arising from the requirements for the validity of the EFT which now read
\begin{equation}
	\epsilon_1 \epsilon_2\ll1 \, , \quad  \text{and} \quad \Omega \epsilon_2 \ll 1 \ . \label{eq:eftGal}
\end{equation}
The validity of the WKB approximation and the above EFT requirements imply that  $\epsilon_2\ll\Omega \ll 1/ \epsilon_2$. In order to have a well-defined $\epsilon$ expansion, we require slightly tighter lower bounds given by $\sqrt{\epsilon_2}\ll\Omega$. While in the shift-symmetric case we had $\epsilon_1\sim\epsilon_2$, here $\epsilon_1$ can in principle be larger since, thanks to the Galileon symmetry, all operators are always suppressed by some power of $\epsilon_2$. The LO or $\mathcal{O}(\epsilon^4)$ corrections simply include the $\epsilon_1^2 \epsilon_2^2$ terms.

\begin{figure}[!ht]
	\begin{center}
		\includegraphics[width=0.5 \textwidth]{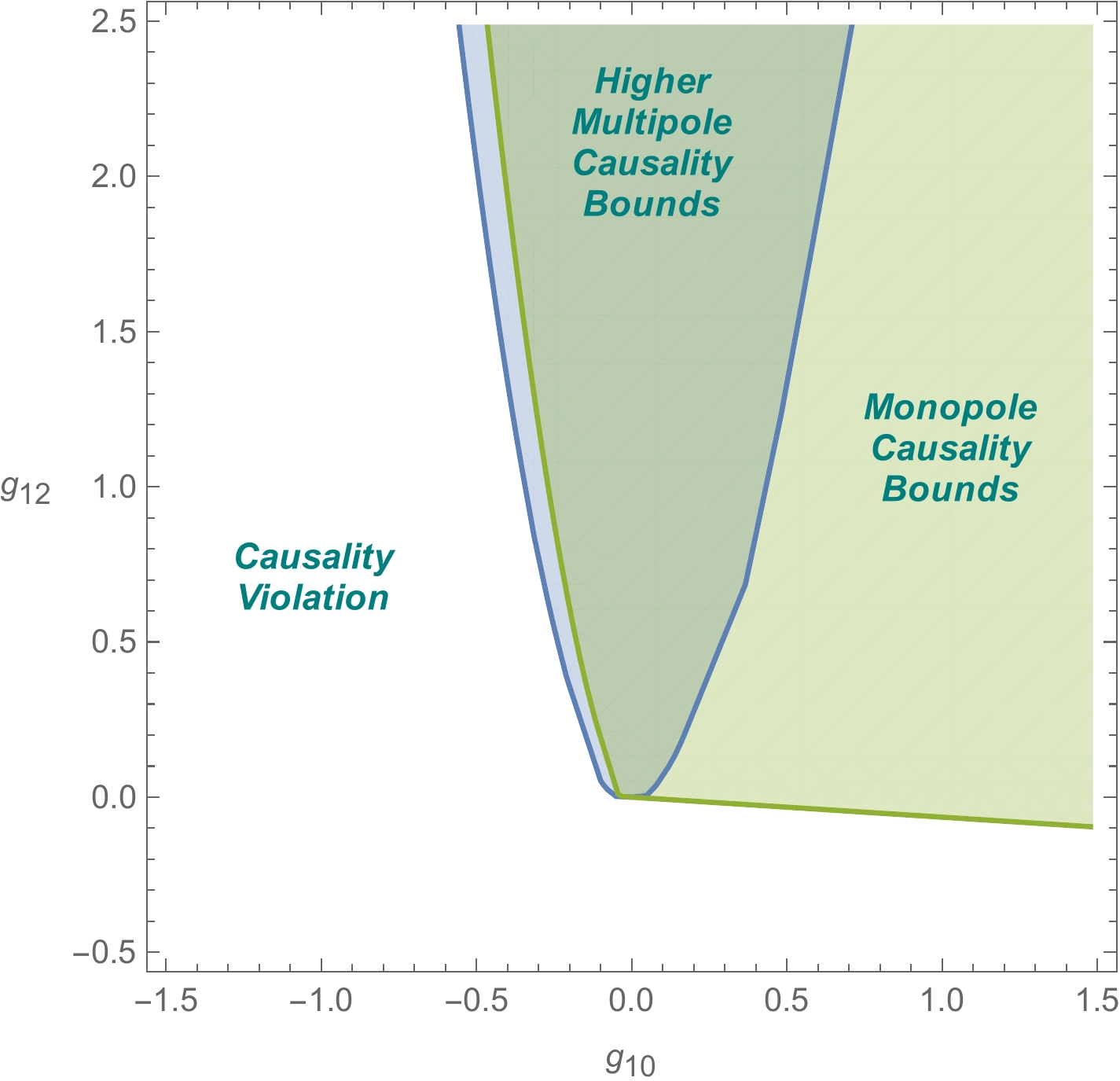}
	\end{center}
	\caption{Causality bounds for the Galileon EFT ($g_8=0$). The green region represents the monopole causality bounds (for the profile considered in Eq.~\eqref{eq:profilefR2}). The blue region represents causality bounds from higher multipole, leading to two-sided bounds. Only the intersection of the blue and green regions is so far causally viable.}
	\label{fig:boundGal}
\end{figure}

As in the previous case, we consider propagation around the background profile in Eq.~\eqref{eq:profilefR2}. When computing the time delay for $\ell=0$ modes, we require that the NLO result differs from the LO only by a $3\%$ for $g_{14}$ of order $1$. As in the shift-symmetric case, we can only get lower bounds on $g_{10}$ in this regime. Note that the monopole constraint for the Galileon symmetry gives $g_{10} \gtrsim 0$, which is nearly as good as it can be for a one-sided bound. Meanwhile, in the higher multipoles case, \ie $\ell>0$, we only consider the leading-order WKB results as in the previous case. For this case, we closely reproduce the $\ell=0$ left-sided bound and get a new maximal right-sided bound as seen in blue in Fig.~\ref{fig:boundGal}.

\section{Discussion and conclusions} \label{sec:Concl}

\begin{figure}[!h]
	\begin{center}
		\includegraphics[width=0.45 \textwidth]{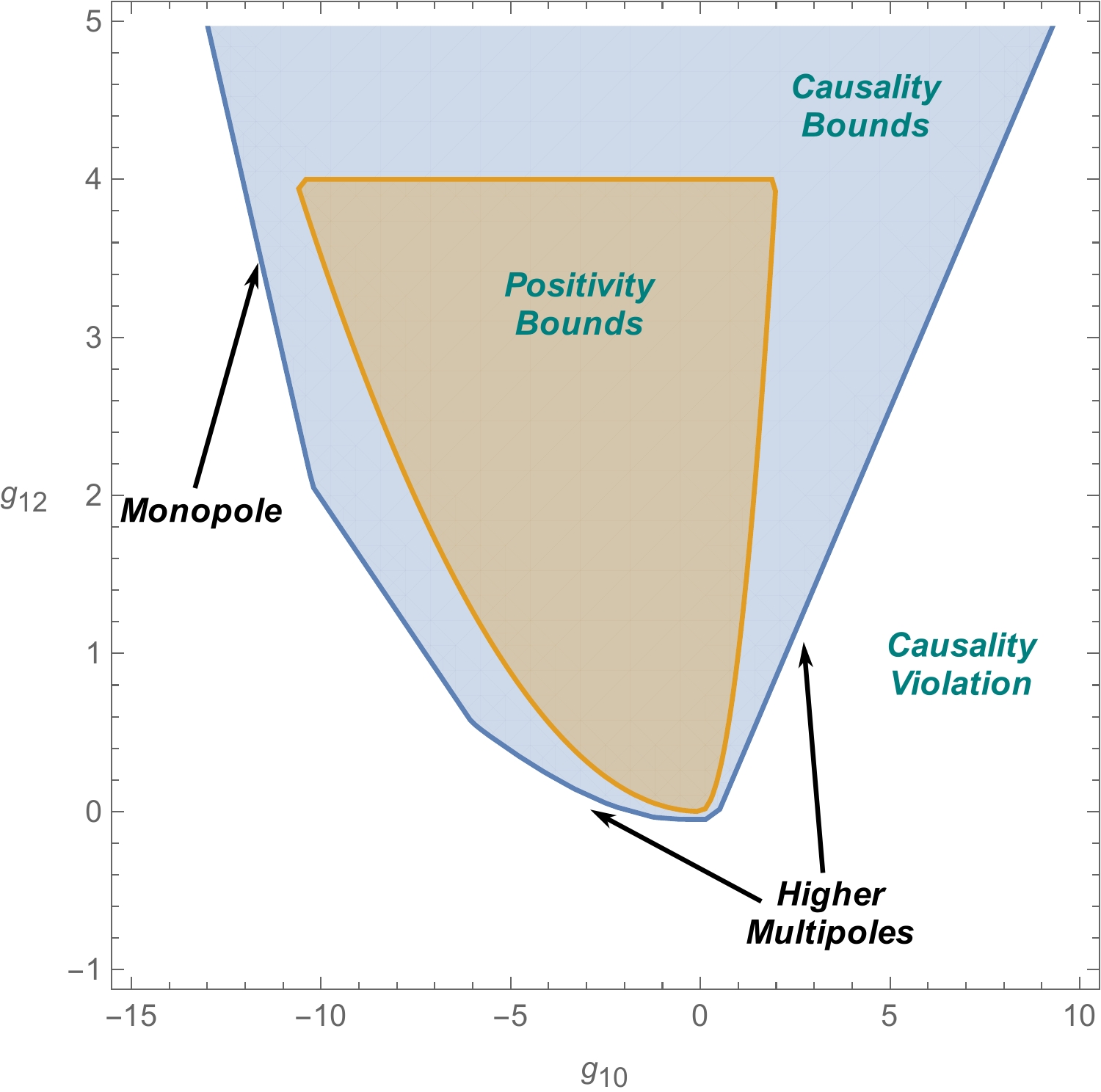}
		\hspace{0.05\textwidth}
		\includegraphics[width=0.45 \textwidth]{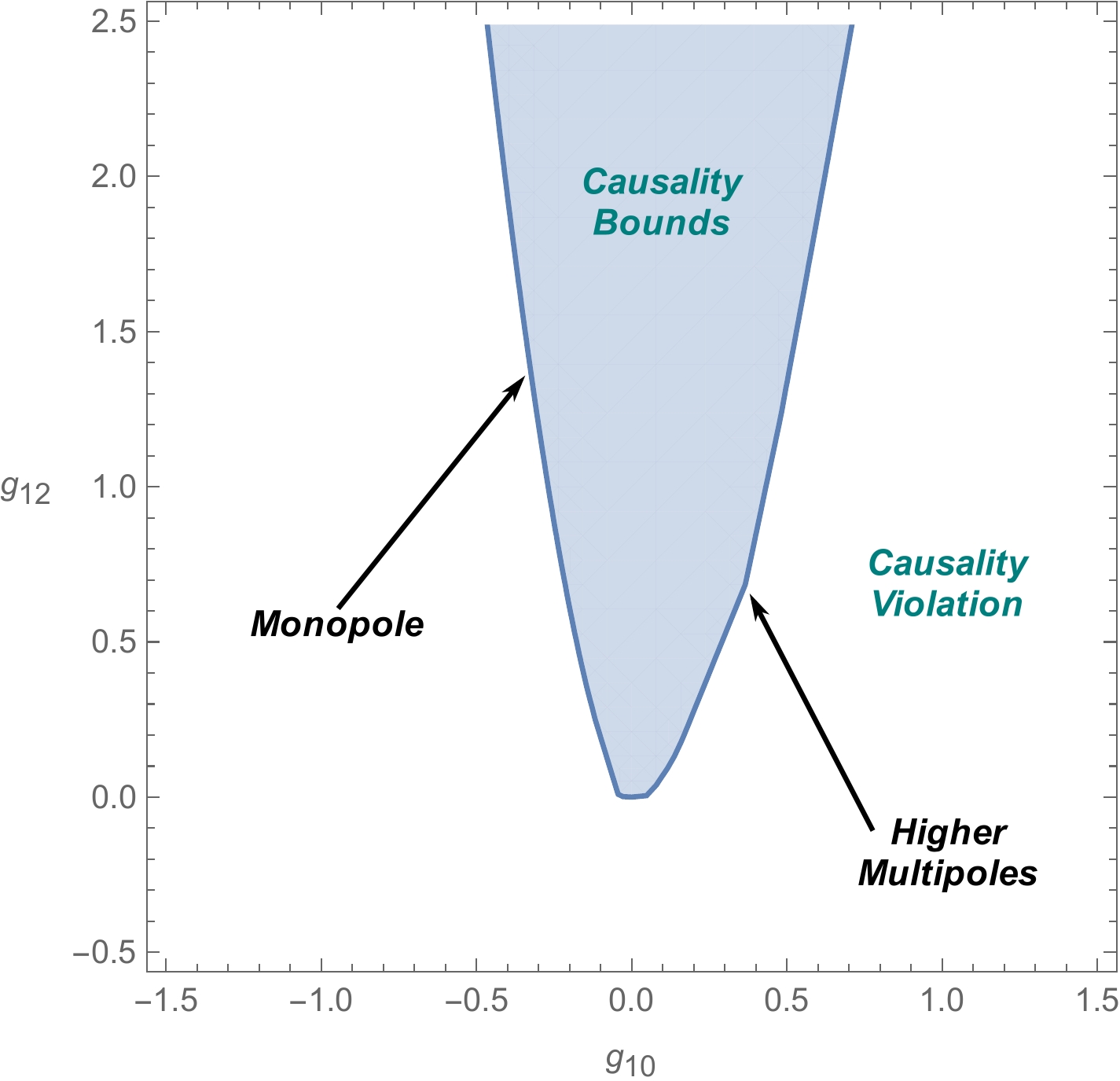}
	\end{center}
	\caption{Infrared Causality constraints on the Wilson coefficients of two scalar low-energy EFT, a shift-symmetric one with $g_8=1$ on the left and a Galileon-symmetric one with $g_8=0$ on the right. In both cases, the white areas are regions in the Wilson coefficients space where a violation of causality can be observed at low-energy,  whereas the orange one is derived from positivity bounds requiring assumptions in the UV.  To obtain these results, we combined lower and upper bounds derived respectively in the $\ell=0$ and $\ell>0$ cases.}
	\label{fig:bounds}
\end{figure}

We have seen that requiring that the effective field theory only leads to causal propagation around a given spherically-symmetric background allows us to put tight bounds on the Wilson coefficients of a low energy EFT, independently of its ultimate high energy completion. Remarkably, there are two physical regimes that give rise to different bounds. The propagation of zero angular momentum partial waves gives rise to lower bounds while the propagation of high $\ell$ modes imposes both lower and upper bounds, although the lower bounds are in general not competitive with those arising from $\ell=0$ modes. We can summarise our findings by combining both results from the monopole and the higher-order multipoles. This is shown in the blue causal regions depicted in Fig.~\ref{fig:bounds}.\\

On the left pane of Fig.~\ref{fig:bounds} we observe the causality bounds (blue) compared to the positivity bounds (orange). While our causality bounds are not as constraining as the positivity ones, we note two important points. First, contrary to the positivity bounds, causality bounds do not require any assumptions of the UV completion (including notably, unitarity and locality) they arise purely from infrared physics that is well described by the EFT. Second, positivity bounds have by now been optimised using various techniques allowing to probe features of the EFT beyond its forward limit, while ours were so far obtained using a simple static and spherically symmetric profile with a simple extremisation procedure. It is likely that tighter bounds could be derived by allowing for more generic and less symmetric profiles.\\

More importantly, we highlight that the precise numerical values of the causal bounds should not be the main focus of our results. The fact that by simply requiring causal propagation in the infrared we can obtain such semi-compact bounds is in itself remarkable.  A naive version of the right-sided positivity bound is given by $g_{10}<2g_8$  and can be derived simply using the $s \leftrightarrow u$ dispersion relation \cite{Tolley:2020gtv}. This bound is slightly optimised when using triple crossing symmetry $s \leftrightarrow t \leftrightarrow u$. Note that in our causality bounds, we only produce an upper bound for $g_{10}$ and lower bound for $g_{12}$ when looking at higher multipoles. On the other hand, the left-sided positivity bounds are fully coming from triple crossing symmetry. In our analysis this lower bound can be reproduced by looking at both high $\ell$ and $\ell=0$ scattering, but the stronger bound comes from the monopole bound. This suggests that our analysis approximately reproduces bounds purely from $s \leftrightarrow u$ dispersion relation in the UV when looking at higher multipoles and triple crossing symmetry when looking at the monopole. However, this seems to be the opposite behaviour of the one observed in \cite{Tolley:2020gtv,Caron-Huot:2020cmc}, where the upper bound is obtained at $\ell=0$ and the lower one at $\ell \geq 2$.\\

Correspondingly, in the right pane  of Fig.~\ref{fig:bounds} we see that requiring infrared causality of the Galileon theory allows us to recover a very similar result to the recently derived full-crossing symmetric positivity bounds that entirely rule out the quartic Galileon by assuming properties of the UV completion. Thus, we effectively rule out the quartic Galileon as a causal low energy scalar effective field theory with no other light degrees of freedom. This does not imply that we rule out the quartic Galileon coupling that would arise in a gravitational setting.  For example, the Galileon theory is a meaningful decoupling limit of massive gravity theories, but can never be considered as a low energy description without the inclusion of other modes. Moreover the Galileon field would generically couple to the trace of the stress-energy tensor, which must obey some consistency conditions of its own. We discuss this point in Appendix~\ref{app:Matter_coupling} and leave for future work the analysis of the situation where we have a gravitational coupling in which one has to impose conditions on the sources to be physical. Instead, our analysis holds if we assume that we are dealing with a  scalar EFT in its own right that can be coupled to an arbitrary external source so that causal propagation is required for any possible external source configuration. \\

Over the past few years, remarkable progress has been made in deriving new sets of non-linear, compact positivity bounds that make use of full $s\leftrightarrow t \leftrightarrow u$ crossing symmetry. This work serves as proof of principle that low energy causality arguments alone can go a long way in making contact with known positivity bounds. This extends the earlier observation of \cite{Adams:2006sv} (for a more recent discussion connecting time delays and positivity bounds see Appendix A of \cite{Arkani-Hamed:2020blm}).
It would be interesting to understand how constraining low energy causality is when optimising the bounds derived in this paper across more general backgrounds similar to that considered in \cite{deRham:2020zyh,Chen:2021bvg,deRham:2021bll}. One might expect that fewer symmetries could lead to stronger bounds. Similarly, one could use this approach to constrain Wilson coefficients of higher derivative terms that arise in the EFT which have been previously bounded using positivity arguments. One appeal of these constraints is that they can easily be generalizable to include operators that are higher order in the field and hence would not contribute at tree-level to known $2\to 2$ positivity bounds.  Furthermore, the requirement of low energy causality can be imposed on gravitational theories and curved backgrounds without running into problems related to the lack of an S-matrix or broken Lorentz symmetries, which would make them particularly appealing for instance for  cosmological \cite{Melville:2019wyy,deRham:2021fpu,Grall:2021xxm} or black hole gravitational bounds \cite{Chen:2021bvg,deRham:2021bll}. In future work, we will explore how causality can give rise to bounds in such situations.

\section{Acknowledgments}
We would like to thank Andrei Khmelnitsky for collaborations in the earlier stages of this work. We would also like to thank Cliff Burgess, Massimo Porrati and the organizers and attendees of the IAS workshop ``Possible and Impossible in Effective Field Theory: From the S-Matrix to the Swampland" for useful discussions. 
The work of MCG, AJT and CdR is supported by STFC grant ST/T000791/1.
 MCG and CdR are supported by the European Union's Horizon 2020 Research Council grant 724659 MassiveCosmo ERC–2016–COG. VP is funded by the Imperial College President's Fellowship.
 CdR thanks the Royal Society for support at ICL through a Wolfson Research Merit Award.  CdR is also supported by a Simons Foundation award ID 555326 under the Simons Foundation Origins of the Universe initiative, Cosmology Beyond Einstein's Theory and by a Simons Investigator award 690508. AJT thanks the Royal Society for support at ICL through a Wolfson Research Merit Award.

\appendix

\section{Causal time advances and Lorentz invariant UV completions}
\label{timeadvance}

As noted by Wigner and Eisenbud \cite{Eisenbud,Wigner:1955zz}, for scattering in a potential of finite range $a$, it is natural to obtain a scattering time advance of $2a/v$ for spherical wave scattering since this reflects the time advance that a wave which scatters directly off the hard boundary at $r=a$, relative to a wave which makes it to $r=0$. Clearly this does not violate causality, and so the causality condition of Wigner-Eisenbud for monopole ($\ell=0$) scattering is
\be\label{WEbound}
\Delta T > - \frac{2a}{v}   -\frac{{\cal O}(1)}{\omega} \, ,
\ee
with $v$ the group velocity of the wave. Given this, one may wonder whether we have been too strict in our consideration of monopole scattering by not allowing any time advance. The key difference is that we are interested in the scattering of essentially massless particles in the relativistic limit for which $\omega $ is large in comparison to the potential $V$, and the scale of variations of the potential $r_0$. More precisely we assume ${\rm Max}[V^{(n)}(r)]\ll \omega^{n+1}$ for all $n \ge 0$.
In this limit, no resolvable time advance is consistent with Lorentz invariant causality. \\

To understand why this is the case, let us consider the case of relativistic scattering off of a \mbox{(quasi-)hard} sphere. To make comparison with the non-relativistic problem, consider a complex massive scalar field $\Phi$ of mass $m$, which is charged under a $U(1)$ gauge field whose Coulomb potential $q A_0 = V(r)$ takes the form
\be
V(r) = V_0 \theta(a-r) \, .
\ee
The equation of motion for the complex scalar is
\be
m^2 \Phi-\nabla^2 \Phi + D_t^2 \Phi =0  \, ,
\ee
where $D_t =\partial_t + i V$. For a given frequency and multipole we have
\be
(\omega-V(r))^2 \Phi = m^2 \Phi-\frac{1}{r^2} \frac{\partial }{\partial r} \( r^2 \frac{\partial \Phi}{\partial r} \) +\frac{\ell(\ell+1)}{r^2} \Phi \, .
\ee
The non-relativistic problem is obtained as usual by replacing $\omega = m+\omega_{\rm NR}$ and neglecting $\omega_{\rm NR}^2$ and $V^2$ terms.
Focussing on the monopole case $\ell=0$ for simplicity, the solution for $r<a$ which is regular at $r=0$ is
\be
\Phi(r) = \frac{A}{r} \sin \(\kappa_0 \,  r\) \, ,
\ee
with $\kappa_0= \sqrt{(\omega-V_0)^2-m^2}$. Denoting $k = \sqrt{\omega^2-m^2}$,
the solution for $r>a$ can be parametrised as
\be
\Phi(r) = \frac{A'}{2i r} \(e^{2i \delta} e^{i k r} -e^{-ik r}  \)   \, .
\ee
Matching at $r=a$ determines the relativistic phase shift to be
\be
e^{2 i \delta } = e^{-2i a k} \frac{\kappa_0 \cos(a \kappa_0)+ i k \sin(a \kappa_0)}{\kappa_0 \cos(a \kappa_0)- i k \sin(a \kappa_0)} \, .
\ee
Now in the true hard sphere limit $|V_0| \rightarrow \infty$ for which the field vanishes for $r<a$
the phase shift reduces to
\be
e^{2 i \delta } = e^{-2i a k}\,,
\ee
and as expected this gives the relativistic version of the time advance noted by Wigner and Eisenbud
\be
\Delta T =2 \frac{\p \delta}{\p \omega}= -\frac{2 a}{v}\,,
\ee
with $v= \frac{\d \omega}{\d k}=k/\omega$, and a similar behaviour occurs even at finite $V_0$  consistent with the bound \eqref{WEbound}.\\

Crucially however this effect occurs because the potential is sharper than the frequencies being considered. If we consider rather the situation where the frequencies are large in comparison to the typical scale of variation of the potential, we may use the WKB approximation for which the phase shift will take the approximate form
\be
\delta = \int_0^{\infty} \d r \( \kappa(r)- \sqrt{\omega^2-m^2} \) \, ,
\ee
where now
\be
\kappa(r) = \sqrt{(\omega-V(r))^2-m^2} \, .
\ee
For $\omega>{\rm Max}\left( |V(r)|\right)$ in the massless case $m=0$, the leading WKB correction to the time delay vanishes for $m=0$ since the leading contribution to the phase shift is frequency-independent. The first order correction to the WKB phase shift gives a frequency-dependent term which gives rise to a time-delay
\be
\Delta T \sim \frac{V'(0)}{\omega^3}+\dots
\ee
In the high frequency limit we are working in where $\omega^2\gg |V'(r)|$ this time delay/advance is unresolvable $|\omega \Delta T|\ll 1$ and higher order WKB corrections are similarly negligible. \\

The massive case is slightly more subtle. The leading WKB term gives a correction
\be
\Delta T \approx  \int_0^{\infty} \d r \frac{m^2 (2 \omega-V(r)) V(r)}{\omega^2 (\omega-V(r))^2} \approx \int_0^{\infty} \d r \frac{2 m^2 V(r)}{\omega^3} \, ,
\ee
where in the first step we assumed $m^2\ll((\omega-V(r))^2,\omega^2)$ and in the last step we assumed $\omega \gg {\rm Max}\left( |V(r)|\right)$. At first sight, it looks like we can easily obtain a time advance from a region of negative potential.  However, for the situations considered in the main text, any background configuration can be parametrised by an overall amplitude and scale in terms of a dimensionless function. Similarly consider a potential of the form $V(r)= V_0 f\left(r/r_0\right)\,,$ where $f(x)$ is a dimensionless function. The maximum time advance relative to a freely propagating massive particle we can create in this region is then of order
\be
|\Delta T| \sim \frac{ m^2 V_0 r_0}{\omega^3}  \, .
\ee
By assumption, for the WKB approximation to be valid we need $\omega \gg r_0^{-1}$. Furthermore we have assumed $V_0 \ll \omega$. Thus we have the bound
\be
\omega |\Delta T| \ll m^2r_0^2\,.
\ee
For the theories considered in the paper, we assume the fundamental field is massless and any effective mass generated for fluctuations around a given background solution will be bounded in the sense $m^2 \lesssim{\cal O}(1) r_0^{-2}$, and hence these potential time advances are unresolvable $\omega |\Delta T| \ll 1 $. Thus provided we consider the region $\omega \gg (r_0^{-1}, {\rm Max}\left( |V(r)|\right))$ we do not expect to obtain any resolvable time advance.\\

In summary, although time advances for monopole scattering are allowed in the non-relativistic and low frequency region without contradicting causality, for the scattering of massless or light (in the scale of the background) high frequency scattering is not expected to lead to any resolvable time advance and this is implicit in our use of this criterion in the main text.

\subsection*{Positivity of Lorentz invariant UV completions}

The previous example was particularly trivial since it does not lead to any interesting time delay at high frequencies. To make it more interesting, and to generate a resolvable time delay, consider now a UV theory of two charged scalars, whose fluctuations may be described by one light field $\Phi$ and one heavy field $H$ with mass $M$. Integrating out the heavy scalar will give EFT corrections to the previously considered theory which describe the scattering and will give rise to a time delay.  Focussing on monopole fluctuations, it is natural to rescale $\Phi = \varphi/r$ and $H = h/r$. We will assume the quadratic action for the monopole fluctuations in the UV completion takes the $U(1)$ invariant form
\ba
S &=& \int \d t \int_0^{\infty} \d r  \int \d \Omega \,  \( |D_t \phi|^2 - | \partial_r \phi |^2-m^2 |\phi|^2  +|D_t h|^2 - | \partial_r h |^2- M^2 |h|^2    \right. \\
 &+&\left. \alpha h^*\partial_r \phi + \beta h^* D_t \phi + \alpha^* h \partial_r \phi^* + \beta^* h (D_t\phi)^* \) \, ,\nn
\ea
where we have dropped any mass mixing terms which can be traded for derivative interactions by a field redefinition.
This is manifestly relativistically causal by virtue of the Lorentz invariant two derivative terms which dominate the dynamics at high energy and determine the causal support of the retarded propagators. Integrating out the heavy field gives a low energy effective theory whose cutoff is $\Lambda = M$ and whose full effective action is
\ba
S &=& \int \d t \int_0^{\infty} \d r  \int \d \Omega \,  \Big( |D_t \phi|^2 - | \partial_r \phi |^2-m^2 |\phi|^2 
+  (\alpha \partial_r \phi + \beta D_t  \phi)^* \frac{1}{M^2+D_t^2-\partial_r^2} (\alpha \partial_r \phi+\beta D_t \phi) \Big) \, .\nn
\ea
The effective dispersion relation is
\be
((\omega-V)^2-k_r^2-m^2) ((\omega-V)^2-k_r^2-M^2)- |\alpha  k_r - \beta (\omega-V)|^2 =0 \, .
\ee
Due to the presence of odd powers of $k_r$ in the dispersion relation, the outgoing and ingoing waves have different magnitudes for their momenta $k_r^{\pm}$ and the WKB scattered wave may be parametrised as
\be
\phi =A(r) \( e^{i \int_0^{r} k_r^+ \d r  }- e^{i \int_0^{r}  k_r^- \d r } \) \, .
\ee
which is matched against the asymptotics
\be
\phi = A'  \( e^{2 i \delta} e^{i \sqrt{\omega^2-m^2} r}- e^{-i  \sqrt{\omega^2-m^2} r} \)\,,
\ee
to give the WKB phase shift
\be
\delta =  \int_0^{\infty} \d r \[\frac{1}{2} (k_r^+(r)+k_r^-(r))- \sqrt{\omega^2-m^2}  \] \, .
\ee
In the regime of validity of the low energy EFT, the leading two derivative terms in the effective action are
\be\label{toyEFT}
 S= \int \d t \int_0^{\infty} \d r  \int \d \Omega \,  \( |D_t \phi|^2 - | \partial_r \phi |^2-m^2 |\phi|^2+\frac{1}{M^2} |\alpha \partial_r \phi+\beta D_t\phi|^2 + \dots  \) \, ,
\ee
and the time delay takes the form
\be
\Delta T = \Delta T_{M=\infty}+ \Delta T_{\rm EFT} \, ,
\ee
where $\Delta T_{M=\infty}$ is the delay obtained previously and the leading EFT correction is
\ba  \Delta T_{\rm EFT} &=& 2 \frac{\p}{\p \omega } \int_0^{\infty} \d r \[\frac{1}{2} (k_r^+(r)+k_r^-(r))-\kappa(r) \] \, , \nn \\
&=& \frac{1}{M^2} \frac{\p}{\p \omega }\int_0^{\infty} \d r \[\frac{1}{2\kappa(r)}\left| \alpha \kappa(r)-(\omega-V) \beta\right|^2 +\frac{1}{2\kappa(r)}\left| \alpha \kappa(r)+(\omega-V) \beta\right|^2   \] + \dots \nn \\
 &=& \frac{1}{M^2}\frac{\p}{\p \omega }\ \int_0^{\infty} \d r \[ |\alpha|^2 \kappa+|\beta|^2 \frac{(\omega-V)^2}{\kappa} \] +\dots \nn \\
 &=&  \frac{1}{M^2} \int_0^{\infty} \d r \[ |\alpha|^2 \frac{(\omega-V)}{\kappa}+|\beta|^2 \frac{(\omega-V)}{\kappa}\(1-\frac{m^2}{\kappa^2}\) \] + \dots  \, .
\ea
In the WKB region considered, $\kappa \gg m$, and $\omega \gg {\rm Max}[V(r)]$ and so both terms are manifestly positive.
Since in this example we know the UV completion, we can directly infer the cutoff in $\omega$ of the low energy EFT by asking at what energy scale does the dispersion relation depart from that implied by the two derivative action \eqref{toyEFT}. This is when $(\omega-V) \sim M^2/(|\alpha|+|\beta|)$ and so we infer that the largest time delay calculable within the low energy EFT we could create is bounded by
\be
|\omega \Delta T_{\rm EFT}| \lesssim (|\alpha|+|\beta|) r_0 \, .
\ee
Since the RHS can be made arbitrarily large by increasing $r_0$, remaining in the region of validity of the low energy EFT, this positive time delay can be made resolvable.
Thus as anticipated, a consistent unitary Lorentz invariant UV completion of an EFT for a massless or light field gives rise to a positive, generally resolvable, time delay $\Delta T>0$ in the WKB region, and the EFT contribution itself is by itself positive $ \Delta T_{\rm EFT}>0$.

\section{Conventions} \label{ap:DefDimLess}
In this Appendix, we summarise some our relations and conventions.
For completeness, we consider the EFT including up to dimension-14 operators and work with the following form of the Lagrangian,
\begin{align}
	\L =& - \frac12 (\p \phi)^2 - \frac12 m^2 \phi^2+ \frac{g_8}{\Lambda^4} (\p \phi)^4 \nonumber \\
	&  + \frac{g_{10}}{\Lambda^6} (\p \phi)^2 \Big[ (\phi_{, \mu \nu})^2  -  (\Box \phi)^2 \Big] + \frac{g_{12}}{\Lambda^8} (( \phi_{, \mu \nu} )^2 )^2 + \frac{g_{14}}{\Lambda^{10}} ( \phi_{, \mu \nu} )^2 ( \phi_{, \alpha \beta \gamma} )^2 \, .
	\label{eq:Lhigh}
\end{align}
The dimension-14 operator is constrained by the following positivity bounds
\begin{equation}
	 -2 g_{12} < g_{14} < \frac{27}{5} (2g_8 - g_{10}) \ .
\end{equation}
The relations between the parameters considered here and those included in \cite{Tolley:2020gtv} and \cite{Caron-Huot:2020cmc} are given in the  Table~\eqref{tab:dictionary} below.\\

\begin{table}[!h]
	\begin{center}
	\begin{tabular}{ | c | c | c | c | } \hline
		 EFT & Tolley \emph{et al.},  \cite{Tolley:2020gtv} & Caron-Huot \emph{et al.}, \cite{Caron-Huot:2020cmc} \\ \hline
		 $g_8$ & $\frac{1}{4}\tilde{a}_{1,0}$ & $\frac{1}{2} \Lambda^4 g_2$ \\
		 $g_{10}$ & $-\frac{1}{3} \tilde{a}_{0,1}$ & $\frac{1}{3} \Lambda^6 g_3$ \\
		 $g_{12}$ & $\tilde{a}_{2,0}$ & $4 \Lambda^8 g_4$ \\
		 $g_{14}$ & $\frac{4}{5}\tilde{a}_{1,1}$ & $-\frac{8}{5}\Lambda^{10} g_5$ \\ \hline
	\end{tabular}
	\caption{Parameters dictionary relating the conventions used in this work, defined in Eq.~\eqref{eq:Lhigh} and others presented in the literature.}
	\label{tab:dictionary}
	\end{center}
\end{table}

In order to extremise the causality bounds, it is convenient to work with dimensionless parameters.
The relations between the dimensionless parameters and their dimensionfull counterparts is provided  in Table \ref{tab:dictionaryDimensionless} below.
\begin{table}[!h]
	\begin{center}
	\begin{tabular}{ | c | c | c | c | c | c | c | c | } \hline
		 Dimensionless parameter & $f(r)$ & $R$ & $\epsilon_1$ & $\epsilon_2$ & $\Omega$ & $B$ & $R_t$ \\ \hline
 &  & &  & &  &  &  \\[-6pt]
		 Definition & $\frac{\bar{\phi}(r)}{\bar{\Phi}_0}$ & $\frac{r}{r_0}$ & $\frac{\bar{\Phi}_0}{r_0 \Lambda^2}$ & $\frac{1}{r_0 \Lambda}$ & $\frac{\omega}{\Lambda}$ & $\frac{b}{r_0}$ & $\frac{r_t}{r_0}$ \\[5pt] \hline
	\end{tabular}
	\caption{Parameters dictionary relating the dimensionless and dimensionfull ones.}
	\label{tab:dictionaryDimensionless}
	\end{center}
\end{table}
It is worth noting that $\bar{\Phi}_0$ carries the scale of the background field $\bar{\phi}$, $r_0$ is its typical scale of variation, whereas $\omega$ is the frequency of the scattered perturbation. The cutoff of the scalar EFT in Eq.~\eqref{eq:L} is given by $\Lambda$ if the dimensionless couplings $g_i$ are all considered to be at most of order 1. Finally, $b$ and $r_t$ are respectively the impact parameter of the free theory and the turning point of the higher-multipole scattering events.

\section{NLO corrections to the time delay at $\ell=0$} \label{ap:NLO}
In this Appendix, we provide the explicit expressions required for computing the time delay at the next order in the EFT, which we refer to as next-to-leading order (NLO).
At NLO, the equation of motion for the monopole $\ell=0$ is given by,
\begin{align}
	&\left. \hat{W}_0(R) \right|_{\rm NLO}= 1152 g_8^3 \epsilon _1^6 f'(R)^6 +224 g_8 g_{12} \Omega ^2 \epsilon _1^4 \epsilon _2^2 f'(R)^2 f''(R)^2  \\
	&+144 \frac{g_8^2}{\Omega ^2} \epsilon _1^4 \epsilon _2^2 \left(2 \frac{f'(R)^3 f''(R)}{R}+2 f'(R)^2 f''(R)^2+f^{(3)}(R) f'(R)^3\right) \nonumber \\
	&-96 g_8 g_{10} \epsilon _1^4 \epsilon _2^2 \left(\frac{f'(R)^4}{R^2}-3\frac{f'(R)^3 f''(R)}{R}\right) \nonumber \\
	&-8 g_{12} \epsilon _1^2 \epsilon _2^4 \left(2\frac{f'(R)^2}{R^4}+\p_R\(4 \frac{f'(R)f''(R)}{R^2} -2 \frac{f''(R)^2}{R} + f^{(3)}(R) f''(R)\)\right) \nonumber \\
	&-4 g_{14} \Omega ^2 \epsilon _1^2 \epsilon _2^4 \left(12\frac{f'(R) f''(R)}{R^3}+4\frac{f^{(3)}(R) f'(R)-3 f''(R)^2}{R^2}+2\frac{f^{(3)}(R) f''(R)}{R}+\p_R\(f^{(3)}(R) f''(R)\)\right) \nonumber \\
	&- 12 \frac{g_{10}}{\Omega ^2} \epsilon _1^2 \epsilon _2^4 \left(\frac{f'(R)^2}{R^4}+\p_R\(\frac{f'(R)f''(R)}{R^2}\)\right) \, .\nonumber
\end{align}
The sound speed square and effective potential are given by
\begin{align}
	&\left. c_s^2(\omega^2,R) \right|_{\rm NLO}= -128 g_8^3 \epsilon_1^6 f'(R)^6 - 96 g_8 g_{12} \epsilon_1^4 \epsilon_2^2 \frac{\omega^2}{\Lambda^2} f'(R)^2 f''(R)^2  \\
	& + 96 g_8 g_{10} \epsilon_1^4 \epsilon_2^2 \left( \frac{f'(R)^4}{R^2} + \frac{f'(R)^3 f''(R)}{R} \right) \nonumber \\
	&+ 8 g_{12} \epsilon_1^2 \epsilon_2^4 \left( 2 \frac{f'(R)^2}{R^4} + \p_R\(4 \frac{f'(R)f''(R)}{R^2} -2 \frac{f''(R)^2}{R} + f^{(3)}(R) f''(R)\) \right) \nonumber \\
	&+ 4 g_{14} \epsilon_1^2 \epsilon_2^4 \frac{\omega^2}{\Lambda^2} \left( 12 \frac{f'(R) f''(R)}{R^3} +4 \frac{-3 f''(R)^2 + f'(R) f^{(3)}(R)}{R^2} +2 \frac{f''(R) f^{(3)}(R)}{R} + \p_R\(f^{(3)}(R) f''(R)\) \right) \ \, , \nonumber
\end{align}
and
\begin{align}
	&\left. V_{\text{eff}}(R) \right|_{\rm NLO}= -48 g_8^2 \epsilon _1^4 \left(2\frac{f'(R)^3 f''(R)}{R}+4 f'(R)^2 f''(R)^2+f^{(3)}(R) f'(R)^3 \right)  \\
	&+12 g_{10} \epsilon _2^2 \epsilon _1^2 \left(\frac{f'(R)^2}{R^4}+\p_R\(\frac{f'(R)f''(R)}{R^2}\)\right) \, . \nonumber
\end{align}
The integrand of the time delay at NLO is given by
\begin{align}
	\left. \mathcal{I}_0(\omega^2,R) \right|_{\rm NLO}= &8 (\omega r_0) \epsilon_1^2 \left[ 104 g_8^3 \epsilon _1^4 f'(R)^6 -2 \frac{g_8^2}{\Omega ^2} \epsilon _1^2 \epsilon _2^2 \left(3 \frac{f'(R)^4}{R^2}-4 f'(R)^2 f''(R)^2\right) - 6 g_8 g_{10} \epsilon _1^2 \epsilon _2^2 \frac{f'(R)^4}{R^2} \right. \nonumber \\
	&\left. +72 g_8 g_{12} \Omega ^2 \epsilon _1^2 \epsilon _2^2 f'(R)^2 f''(R)^2 -\frac{45}{4} g_{14} \Omega ^2 \epsilon _2^4 \left(\frac{f'(R)^2}{R^4} + \frac{f^{(3)}(R) f'(R)-f''(R)^2}{R^2}\right) \right. \nonumber \\
	&\left. + \left(g_{12}-\frac34 \frac{g_{10}}{\Omega ^2}\right) \epsilon _2^4 \left(\frac{f'(R)^2}{R^4}-\frac{f^{(3)}(R)f'(R)+f''(R)^2}{R^2}\right) \right] \nonumber \\
	&+ \text{total derivatives}  \, .
	\label{eq:I0NLO}
\end{align}
We do not write the total derivative terms explicitly since they vanish upon integration in the $\ell=0$ case considered here. Note that the total derivatives include terms like $f'(R)^2/R^3, f'(R)f''(R)/R^2$ and $f''(R)^2/R$ that diverge when evaluated at the origin. In this analysis, we have been careful to cancel the divergences so that the total derivatives in the last line of Eq.~\eqref{eq:I0NLO} actually vanish upon integration from $0$ to $\infty$.

\section{Higher-order multipoles} \label{ap:NLOmultipole}

In this Appendix, we provide the leading-order expressions to the various functions entering the computation of the time delay for $\ell>0$, as defined in Section \ref{sec:multipole}. Note that since we are focusing on a regime where $\omega r_0 \gg 1$ in order to safely ignore all WKB corrections, we will ignore all $1/\Omega$ corrections for consistency. Furthermore, such a regime also allows us to forget about NLO corrections, hence they will be omitted here. The function $W_{\ell}(R)$ reads, at leading order,

\begin{align}
	\left. W_{\ell}(R) \right|_{\rm LO} = &\left(1-\frac{B^2}{R^2}\right) \left(1+ 8 g_8 \epsilon _1^2 f'(R)^2 + 96 g_8^2 \epsilon _1^4 f'(R)^4 \right)  \\
	&+ 8 g_{12} \Omega ^2 \epsilon _1^2 \epsilon _2^2 \left\lbrace \left(1-\frac{B^2}{R^2}\right) \left(f''(R)-\frac{f'(R)}{R}\right)+\frac{f'(R)}{R}\right\rbrace^2 \nonumber \\
	& +12 g_{10} \epsilon _1^2 \epsilon _2^2 \frac{B^2}{R^2} \left(\frac{f'(R)^2}{R^2}-\frac{f'(R)f''(R)}{R}\right) \nonumber \,.
\end{align}
This means that the square sound velocity and the effective potential at leading order read

\begin{align}
	\left. c_s^2(\omega^2,R) \right|_{\rm LO}=& 1 -8 g_8 \epsilon _1^2 f'(R)^2 -32 g_8^2 \epsilon _1^4 f'(R)^4  \\
	& -8 g_{12} \Omega ^2 \epsilon _1^2 \epsilon _2^2 \left(2\frac{B^2}{R^2} \left(\frac{f'(R)f''(R)}{R}-f''(R)^2\right)+f''(R)^2\right) -24 g_{10} \epsilon _1^2 \epsilon_2^2 \frac{f'(R) f''(R)}{R} \, ,\nonumber \\
	\label{eq:velocityHighL}
	\left. V_{\text{eff}}(R) \right|_{\rm LO}=& \frac{L^2}{R^2} \left[ \vphantom{\frac12} 1 -8 g_{12} \Omega ^2 \epsilon_1^2 \epsilon_2^2 f''(R)^2 -12 g_{10} \epsilon_1^2 \epsilon_2^2  \left(\frac{f'(R)^2}{R^2}+\frac{f'(R) f''(R)}{R}\right) \right]  \\
	&-8\frac{L^4}{R^4} g_{12} \epsilon _1^2 \epsilon _2^4 \left(\frac{f'(R)^2}{R^2}-f''(R)^2\right) \, .\nonumber
\end{align}
We have decided to express the effective potential in terms of the orbital number $L$ rather than the reduced effective impact parameter $B$ in order to make contact with the free theory where $V_{\rm eff, free}= L^2/R^2$. It is worth mentioning once again that the effective potential term is suppressed by $(\omega r_0)^{-2}=\epsilon_2^2 / \Omega^2$ with respect to the speed of sound term. Hence, the leading-order effective potential should include terms up to $\mathcal{O}(\epsilon_1^2)$. Note that the terms $L^2 \epsilon_1^2 \epsilon_2^2$ and $L^4 \epsilon_1^2 \epsilon_2^4$ seem to be higher-order and appear to be unnecessarily taken into account. However, recalling that $L=\Omega B/\epsilon_2$ implies $\epsilon_2 L \sim \mathcal{O}(\epsilon^0)$. This means that $L^2 \epsilon_1^2 \epsilon_2^2 \sim L^4 \epsilon_1^2 \epsilon2_4 \sim \mathcal{O}(\epsilon_1^2)$, so all terms considered are indeed leading order in the effective potential. To show that this is indeed the correct functional form for the potential, one could rewrite the term of interest, \ie $V_{\rm eff}/(\omega r_0)^2$ rather than just the effective potential, in terms of the variable $B$ that does not hide any dependence on $\epsilon_i$,
\begin{align}
	\left. \frac{V_{\text{eff}}(R)}{(\omega r_0)^2} \right|_{\rm LO}=& \frac{B^2}{R^2} \left[ \vphantom{\frac12} 1 -8 g_{12} \Omega ^2 \epsilon_1^2 \epsilon_2^2 f''(R)^2 -12 g_{10} \epsilon_1^2 \epsilon_2^2  \left(\frac{f'(R)^2}{R^2}+\frac{f'(R) f''(R)}{R}\right) \right] \\
	& - 8 \frac{B^4}{R^4} g_{12} \Omega ^2 \epsilon _1^2 \epsilon _2^2\left(\frac{f'(R)^2}{R^2}-f''(R)^2\right)  \, . \nonumber
\end{align}
In this set up, the corresponding turning point is now $R_t$, which is given by
\begin{equation}
	\left. R_t \right|_{\rm LO} = B \left[ 1 - 4 g_{12} \Omega^2 \epsilon_1^2 \epsilon_2^2 \frac{f'(B)^2}{B^2} - 6 g_{10} \epsilon_1^2 \epsilon_2^2 \left(\frac{f'(B)^2}{B^2}+\frac{f'(B) f''(B)}{B}\right) \right] \,.
\end{equation}
Moreover, we have
\begin{align}
	&\left. U_{\ell}(R) \right|_{\rm LO} = 4 \left(1-\frac{B^2}{R^2}\right) \left( g_8 \epsilon _1^2 f'(R)^2 + 10 g_8^2 \epsilon _1^4 f'(R)^4 \right)  \\
	& -6 g_{10} \epsilon _1^2 \epsilon _2^2 \left\lbrace \left(1-\frac{B^2}{R^2}\right) \left(\frac{f'(R)^2}{R^2}-\frac{f'(R) f''(R)}{R}\right)- \left( \frac{f'(R)^2}{R^2} + \frac{f'(R) f''(R)}{R} \right) + \frac{B^2}{R^2} \left( \frac{f'(B)^2}{B^2} + \frac{f'(B) f''(B)}{B} \right) \right\rbrace \nonumber \\
	& +4 g_{12} \Omega^2 \epsilon _1^2 \epsilon _2^2 \left\lbrace \left[\left(1-\frac{B^2}{R^2}\right) \(\frac{f'(R)}{R}-f''(R)\)-\frac{f'(R)}{R}\right]^2 - \frac{B^2}{R^2}\frac{f'(B)^2}{B^2} \right\rbrace \,,\nonumber
\end{align}
with $U_{\ell}(R_t)=0$. Having all the ingredients, the dimensionless time delay can now be expressed in the following form
\ba
\omega \Delta T_{b}(\omega) = (\omega r_0) \left[ \int_B^{\infty} \left( \frac{\Upsilon^{(0)}_{\ell}(R)}{\sqrt{ 1 - \frac{B^2}{R^2} }} + \Upsilon^{(1)}_{\ell}(R) \sqrt{ 1 - \frac{B^2}{R^2} } + \Upsilon_{\ell}^{(2)} \left( 1 - \frac{B^2}{R^2} \right)^{3/2} \right) \mathrm{d}R + \Upsilon_{\ell}^{(3)} \right],\
\ea
where
\begin{align}
	\Upsilon_{\ell}^{(0)}(R) =& 12 g_{10} \epsilon _1^2 \epsilon _2^2 \left\lbrace \frac{f'(R)^2}{R^2}+\frac{f'(R)f''(R)}{R} -\frac{B^2}{R^2} \left(\frac{f'(B)^2}{B^2}+\frac{f'(B) f''(B)}{B}\right) \right\rbrace  \\
	& +24  g_{12} \Omega ^2 \epsilon _1^2 \epsilon _2^2 \left(\frac{f'(R)^2}{R^2}- \frac{B^2}{R^2}\frac{f'(B)^2}{B^2}\right) \,,\nonumber \\
	\Upsilon_{\ell}^{(1)}(R) =& 8 g_8 \epsilon _1^2 f'(R)^2+80 g_8^2 \epsilon _1^4f'(R)^4 - 48 g_{12}\Omega ^2 \epsilon _1^2 \epsilon _2^2 \left(\frac{f'(R)^2}{R^2} - \frac{f'(R) f''(R)}{R} \right)  \\
	&-12 g_{10} \epsilon _1^2 \epsilon _2^2 \left(\frac{f'(R)^2}{R^2}-\frac{f'(R) f''(R)}{R}\right) \,, \nonumber\\
	\Upsilon_{\ell}^{(2)}(R) =& 24 g_{12} \Omega ^2 \epsilon _1^2 \epsilon _2^2 \left(\frac{f'(R)^2}{R^2}-\frac{2 f'(R) f''(R)}{R}+f''(R)^2\right) \,, \\
	\Upsilon_{\ell}^{(3)}(R) =& 12 g_{12} \pi B \Omega ^2 \epsilon _1^2 \epsilon _2^2 \frac{f'(B)^2}{B^2} \,.
\end{align}
Note that $\Upsilon_{\ell}^{(0)}(R) = \mathcal{F}[f(R)] - B^2/R^2 \mathcal{F}[f(B)]$, where $\mathcal{F}$ is a functional of the function $f$. This immediately shows that $\Upsilon_{\ell}^{(0)}(R=B)=0$, hence avoiding any divergence around the lower bound of the integral.

\section{Extremisation method} \label{ap:procedure}

The method used to extremise the causality bounds for the simple profile considered in \eqref{eq:profilefR2} (with $p=3$) is summarised below. In principle the same method could be applied to more generic profiles and in less symmetric situations. The dimensionless time delay is given as a function of
\begin{equation}
	(\omega \Delta T) = (\omega \Delta T)(g_{10}, g_{12}, \mathcal{P}) \,
\end{equation}
where the parameters are listed in the vector
\begin{equation}
	\mathcal{P} = \left\{ g_8, a_0, a_2, a_4, a_6, \epsilon_1, \epsilon_2, \Omega, B \right\} \,.
\end{equation}
In our analysis $g_8$ will be fixed to be either $0$ or $1$ but we include it for completeness. In order to remain within the regime of validity of the EFT we only consider $-5<a_i<5$ so that $f(R)$ is $\mathcal{O}(1)$. More importantly, during the extremisation procedure we constrain the parameters in $\mathcal{P}$ such that the analysis remains in the regime of validity of the EFT as given in Eq.~\eqref{eq:eps} (Eq.~\eqref{eq:eftGal} for the Galileons) by replacing $\ll1$ by $<1/2$. Since the suppression of higher-order EFT corrections always comes as the square of these parameters, this ensures that the terms that we neglect are suppressed by at least a factor of $0.25$. Furthermore, we also need to ensure that the WKB formula is valid up to the order that we compute it. To do so, we explicitly compute corrections to the WKB formula in the monopole case and check that they are negligible. For higher multipoles however, we rely instead on dimensional analysis to compute the order of magnitude of the corrections that are being neglected. This requires enforcing Eq.~\eqref{eq:WKBreq}. For a more detailed discussion on the validity of the EFT and WKB approximation we refer to the analysis in Sections \ref{sec:monopole} and \ref{sec:multipole}. Note that in our analysis, we separated the case $\ell=0$ and $\ell>0$, and also $g_8=0$ and $g_8=1$, which gave four separate sets of causal regions. However, the method used in each of them was identical and will be detailed below.\\

The boundary of the causal region for a given set of parameters is defined by $(\omega \Delta T) = -1/2$, which can be solved for $g_{12}$ to give the equation of a line in the $(g_{10},g_{12})$-plane
\begin{equation}
	g_{12} = m(\mathcal{P}) g_{10} + p(\mathcal{P}) \equiv \mathcal{Y}_{\mathcal{P}}(g_{10}) \,.
\end{equation}
Now, the extremisation process differentiates between lower and upper bounds. In both cases, let us define a vector $\mathcal{G}$ corresponding to set of discrete points in the interval $[0,2.5]$. The parameter $g_{12}$ will take values drawn from $\mathcal{G}$, \ie $g_{12} \in \mathcal{G}$.

The tightest lower bound for $g_{10}$ for a given value of $g_{12}=\mathcal{G}_i$ is achieved by finding the optimal set of parameters $\mathcal{P}_i^{(\rm lower)}$ such that the negative value of $g_{10}$ at the intersection between the two lines defined by $g_{12}=\mathcal{Y}_{\mathcal{P}_i^{(\rm lower)}}(g_{10})$ and $g_{12}=\mathcal{G}_i$ is maximal. It can be defined as

\begin{equation}
	\mathcal{P}_i^{(\rm lower)}= {\rm Max} \left\{ g_{10} < 0 \left| g_{12}=\mathcal{Y}_{\mathcal{P}}(g_{10})\, \& \, g_{12}=\mathcal{G}_i \right. \right\} \,,
\end{equation}
and the `causal' region\footnote{This method does not `prove' causality, it simply indicates the absence of obvious acausality.} $\mathcal{R}_i^{(\rm lower)}$ would consist of all points in the $(g_{10}, g_{12})$-plane that are ``above" this line, meaning
\begin{equation}
	\mathcal{R}_i^{(\rm lower)} = \left\{ (g_{10},g_{12}) \left| g_{10} \in \mathbb{R}, g_{12} > \mathcal{Y}_{\mathcal{P}_i^{(\rm lower)}}(g_{10}) \right. \right\} \,.
\end{equation}
Equivalently, the tightest upper bound for a given $i$ is given by
\begin{equation}
	\mathcal{P}_i^{(\rm upper)}= {\rm Min} \left\{ g_{10} > 0 \left| g_{12}=\mathcal{Y}(\mathcal{P})\, \& \, g_{12}=\mathcal{G}_i \right. \right\} \,,
\end{equation}
and the associated `causal' region
\begin{equation}
	\mathcal{R}_i^{(\rm upper)} = \left\{ (g_{10},g_{12}) \left| g_{10} \in \mathbb{R}, g_{12} < \mathcal{Y}_{\mathcal{P}_i^{(\rm upper)}}(g_{10}) \right. \right\} \,.
\end{equation}
Note that in the case where $\ell=0$, the method does not identify any upper bound, as described previously. This process is iterated for all values of $i$ (and it could be optimised further by exploring more values in the range $[0,2.5]$ or by extending this range) and the final causal region $\mathcal{R}_{\rm causal}$ is obtained by taking the union of all lower and upper regions labelled by $i$,
\begin{equation}
	\mathcal{R}_{\rm causal} = \cup_{i} \cup_{j=\text{lower, upper}} \mathcal{R}_i^{(j)} \,.
\end{equation}

\section{Gravitationally-coupled Galileons}
\label{app:Matter_coupling}
In most of this work, we have considered the scalar field EFT to describe a single low energy degree of freedom in its own right in flat spacetime and in the absence of any other light degrees of freedom. For such low energy EFTs, one can in principle consider an arbitrary external source $J$ that would spontaneously generate an arbitrary (Lorentz-violating) background profile for the scalar field.

We now explore a `Galileon' field which, in some contexts, can be thought of as describing a degree of freedom reminiscent of an infrared modification of gravity (as is the case from instance in the Dvali-Porrati-Gabadadze model \cite{Dvali:2000hr} or massive gravity \cite{deRham:2010ik,deRham:2010kj}). In this case the EFT is not precisely a low energy description, and the presence of other light degrees of freedom may not always be safely ignored. Generating a non-trivial profile for the field typically comes at the price of introducing a non-trivial stress-energy tensor which would also be expected to ever-so-slightly affect the geometry. The subtle issue of backreaction on the geometry can be put aside for now, but in this Appendix, we establish which source would be required to generate the spherically-symmetric background profile we have considered so far. In particular we explore whether there are any physical requirements to be imposed on that source, and whether source satisfies the null or weak energy condition.  In the present case, we consider the coupling of the Galileon to matter through the trace of the stress-energy tensor $T^{\mu}_{\phantom{\mu} \mu}$ which generically arises in massive gravity theories. Thus the source in Eq.~\eqref{eq:lag} is now given by
\begin{equation}
	J=\frac{1}{M_{\rm Pl}}  T^{\mu}_{\phantom{\mu}\mu} \ ,
\end{equation}
The Galileon interactions (and possible mass term) are small corrections compared to the kinetic term and thus the equation of motion for the source reads
\begin{equation}
	\Box \phi = - \frac{g_{\rm matter}}{M_{\rm Pl}} T^{\mu}_{\phantom{\mu}\mu} \,.
	\label{eq:boxphiT}
\end{equation}
The stress-energy tensor needs to respect the spherical symmetry and hence can be written in the following form
\begin{equation}
	T^{\mu}_{\phantom{\mu}\nu} = \text{diag}(- \rho(r), p_r(r), p_{\Omega}(r), p_{\Omega}(r)) \,,
\end{equation}
where $p_r$ and $\rho$ are respectively the radial pressure and energy density of the fluid, and $p_{\Omega}$ is the angular pressure. For simplicity, we write $p_{\Omega} = A p_r$, where $A$ is a constant that will be constrained by requiring asymptotic flatness of the spacetime. The trace of the stress-energy tensor is then simply given by $T^{\mu}_{\phantom{\mu}\mu} = p_r (1+2A) - \rho$.   Energy-momentum conservation implies
\begin{equation}
	p'_r + 2(1-A) \frac{p_r}{r} = 0 \,.
\end{equation}
This first-order differential equation for the radial pressure $p_r$ is solved by,

\begin{equation}
	p_r(r) = \bar p_r\  r^{-2(1-A)} \,, \qquad \rho(r) = \bar p_r\  (1+2A) r^{-2(1-A)} - T^{\mu}_{\phantom{\mu}\mu}(r) \,,
\end{equation}
Asymptotically flatness (or `vacuum'), demands that at large radius $p_r, \rho \sim r^n$ with $n<-3$, which effectively provides the bound $A<-1/2$. Furthermore, for the source to be physical, we should at the very least demand the weak energy condition which requires
\begin{equation}
	\rho > 0 \,, \qquad  \rho + p_r > 0 \,, \qquad {\rm and }\qquad \rho+A p_r>0 \,. \label{eq:WEC}
\end{equation}
Defining $T_{\rm max} = {\rm Max}_{r>0} \left\{ r^{2(1-A)} \left| T^{\mu}_{\phantom{\mu}\mu}(r) \right| \right\}$, then if one were to choose

\begin{equation}
	A<-1 \,, \qquad \bar p_r < \frac{T_{\rm max}}{2(1+A)}<0\,,
	\label{eq:solWEC}
\end{equation}
and as long as $\left| T^{\mu}_{\phantom{\mu}\mu}(r) \right|$ is bounded and $r^{2(1-A)} T^{\mu}_{\phantom{\mu}\mu}(r) \rightarrow 0$ when $r \rightarrow \infty$, which is ensured for exponentially suppressed background profiles as the one considered in Eq.~\eqref{eq:profilefR2}, then the weak energy condition is respected. Note that if one is only interested in the null energy condition, then $\rho$ is unconstrained, but to satisfy the other two conditions in Eq.~\eqref{eq:WEC} we still require that Eq.~\eqref{eq:solWEC} holds. We have thus proven that some fluids with negative pressure along some direction (and positive pressure along others) can represent a physical source generating an asymptotically flat spacetime, satisfying  the weak energy condition and leading to any bounded profile $\bar \phi(r)$. Note that this stress-energy tensor diverges at the origin, indicating that the source ought to be regularised but since  the scalar field remains finite, one would not expect the regularization to impact the outcome of this study.
\bibliographystyle{JHEP}
\bibliography{refs}
	
\end{document}